\documentclass[aps,prb,twocolumn]{revtex4-1}

\usepackage{graphicx}
\usepackage{colortbl}
\usepackage{amssymb, latexsym, textcomp,  textcomp}
\usepackage{amsmath}    
\usepackage{hyperref}
\usepackage{physics}
\usepackage{comment}
\usepackage{pgfplots}
\usepackage{tikz}
\usepackage[caption=false]{subfig}

\pgfplotsset{compat=1.11}

\def\*#1{\mathbf{#1}}

\def \be{\begin{equation}}
\def \ee{\end{equation}}

\begin{document}


\title{Strong coupling of two-dimensional electron ensemble to a single-mode cavity resonator}


\author{Jiabao Chen}
\email[]{jiabao.chen@oist.jp}
\author{Oleksiy Zadorozhko}

\author{Denis Konstantinov}
\email[]{denis@oist.jp}
\affiliation{Quantum Dynamics Unit, Okinawa Institute of Science and Technology (OIST) Graduate University, Tancha 1919-1, Okinawa 904-0412, Japan.}


\date{\today}

\begin{abstract}
We investigate the regime of strong coupling of an ensemble of two-dimensional electrons to a single-mode cavity resonator. In particular, we realized such a regime of light-matter interaction by coupling the cyclotron motion of a collection of electrons on the surface of liquid helium to the microwave field in a semi-confocal Fabry-Perot resonator. For the co-rotating component of the microwave field, the strong coupling is pronouncedly manifested by the normal-mode splitting in the spectrum of coupled field-particle motion. We present a complete description of this phenomenon based on classical electrodynamics, as well as show that the full quantum treatment of this problem results in mean-value equations of motion that are equivalent to our classical result. For the counter-rotating component of the microwave field, we observe a strong resonance when the microwave frequency is close to both the cyclotron and cavity frequencies. We show that this surprising effect, which is not expected to occur under the rotating-wave approximation, results from the mixing between two polarization components of the microwave field in our cavity.  
\end{abstract}


\maketitle

\section{Introduction} 
Interest in collective enhancement of light-matter interaction in an $N$-particle system coupled to a single-mode cavity resonator traditionally comes from research in Atomic Physics and Quantum Optics. Of particular interest is the regime of so-called strong coupling, when the rate of energy exchange between particles and a cavity mode, which for a many-particle system is enhanced by a factor of $\sqrt{N}$,~\cite{Dicke1954} exceeds the dissipation rates set by cavity losses and relaxation processes in the many-particle system. In experiment, the strong coupling is manifested by the normal-mode splitting in the spectrum of coupled field-particle motion, with the splitting given by twice the $\sqrt{N}$-enhanced coupling constant~\cite{kaluzny1983observation,raizen1989normal,zhu1990vacuum,thompson1992observation}. It has been mentioned that this splitting is essentially a classical effect, which can be understood on the ground of two coupled damped oscillators~\cite{zhu1990vacuum,carmichael1989subnatural}, and that observation of quantum electrodynamic (QED) features require photon correlation experiments~\cite{carmichael1991quantum}. 

Recently, interest in collective coupling was revived due to its applications in hybrid quantum systems and quantum technologies~\cite{imamouglu2009cavity,soykal2010strong}. Motivated by proposals to use solid-state systems strongly coupled to microwave (MW) resonators for efficient quantum memory storage~\cite{wesenberg2009quantum,diniz2011strongly,julsgaard2013quantum}, a large body of experimental work has been reported using solid-state spin ensembles~\cite{schuster2010high,kubo2010strong,wu2010h,amsuss2011cavity,abe2011electron,huebl2013high,goryachev2014high,tabuchi2014tabuchi,zhang2014strongly,abdurakhimov2015normal}. In these works, the effect of normal-mode splitting is usually accounted for by the cavity QED theory. Significant experimental work has been also reported using other solid-states systems, in particular two-dimensional electron systems (2DESs) in semiconductors.~\cite{dini2003prl,sapienza2008prl,todorov2010prl,todorov2014prb,todorov2015prb,muravev2011prb,muravev2013prb} Of particular interest is the strong coupling of light to the cyclotron motion of 2DES induced by an applied magnetic field, which was first discussed by Shikin in the context of classical electrodynamics.~\cite{shikin2002} Later on, 2DESs in semiconductor quantum wells and graphene were suggested as a good candidate to reach an ultra-strong coupling regime using a QED treatment.\cite{ciuti2005prb,hagenmuller2010ultrastrong,hagenmuller2012prl,chirolli2012prl} This has given rise to several interesting experimental works~\cite{scalari2012ultrastrong,zhang2016collective,li2018bloch}, including most recent observations of the modification of quantum magneto-transport~\cite{bartolo2018prb,paravicini2018arXiv} and softening of polariton modes~\cite{keller2018arXiv} in ultra-strongly coupled light-matter systems. Despite this very significant amount of work, the distinction between classical and full quantum treatments of the problem of strong light-matter interaction hasn't been fully discussed yet.         

Here, we present our study of the strong coupling regime realized in an ultra-clean 2DES formed on the surface of liquid helium. In the experiment described here, which is an extension of our earlier work,~\cite{abdurakhimov2016strong} we couple the cyclotron motion of electrons in a perpendicular magnetic field to the microwave field in a semi-confocal Fabry-Perot resonator. Owing to an enhanced quality factor of the resonator, we are able to resolve interaction of the electron system with two polarization components of the single-mode microwave field. For the co-rotating component, the strong coupling is pronouncedly manifested by the normal-mode splitting in the spectrum of coupled field-particle motion. Unlike most of the theoretical approaches appearing in recent literature,~\cite{hagenmuller2010ultrastrong,scalari2012ultrastrong,zhang2016collective,li2018bloch} we present a complete description of this phenomenon based on classical electrodynamics. To reconcile this result with other theoretical treatments, we show that the full quantum theory applied to this problem results in mean-value equations of motion that are equivalent to our classical result. For the counter-rotating component of the microwave field, we observe a strong resonance when the microwave frequency is close to both the cyclotron and cavity frequencies. We show that this surprising effect, which is not expected to occur under the rotating-wave approximation applicable under the conditions of our experiment, results from the mixing between two circular polarization components of the microwave field in our cavity. Even though this is a completely classical effect, we show that it is convenient to use a full quantum model to reproduce the experimental results.     

In Sections II and III, we provide details of our experiment and our obtained results. In Section IV, we present a model of our experiment based on the classical equations of motion for electromagnetic field that accounts for two independent polarization degrees of freedom. In Section V, we present a full quantum model and compare this to our classical model. Comparison with experimental results is done by introducing the input-output relations. In Section~\ref{additional}, we present explanation for the resonance appearing under the pumping with the counter-rotating component of the microwave field. This paper concludes with a summary of the obtained results.          

\section{Experiment}

\begin{figure}
	\centering
	\includegraphics[width=8cm]{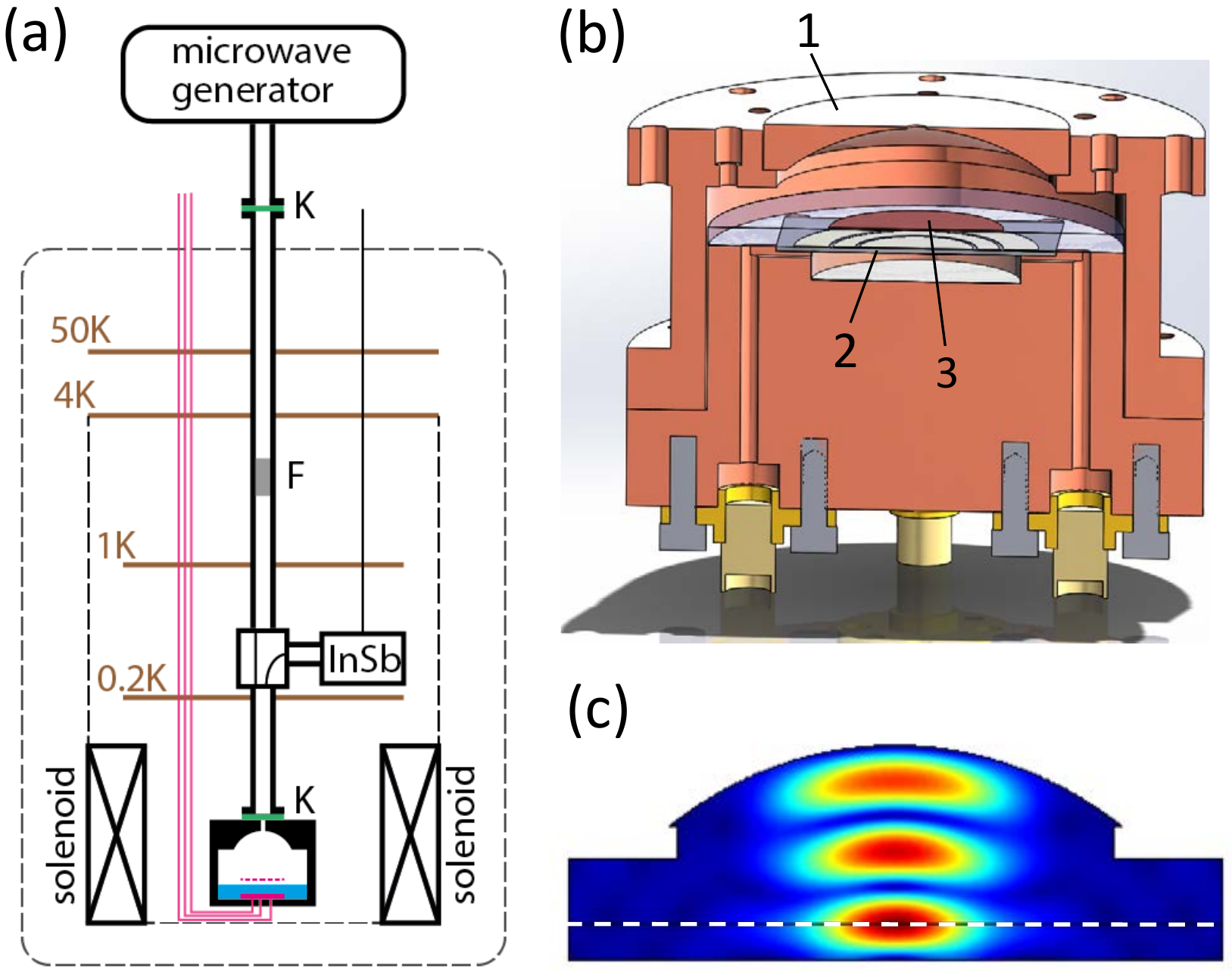}
	\caption{(colr online) (a) Schematic diagram of the experimental setup. (b) 3D drawing of the experimental cell: 1. top spherical mirror, 2. bottom flat mirror with Corbino electrodes; 3. 2DES on the surface of liquid helium. (c) Distribution of the MW electric field of the resonant TEM$_{002}$ mode inside the Fabry-Perot resonator. The dashed (white) line shows the position of the liquid helium level in the resonator and coincides with the position of the first antinode of the MW electric field in the resonator.}
	\label{fig:1}       
\end{figure}

Our experimental method is similar to that described previously~\cite{abdurakhimov2016strong}. A 2DES was created on the surface of superfluid $^4$He cooled to $T=0.2$~K in a vacuum-tight copper cell attached to the mixing chamber of a dilution refrigerator, see Fig.~\ref{fig:1}. The cell contained a semi-confocal Fabry-Perot resonator formed by a top spherical mirror made of copper and a bottom flat mirror made of a 0.5~$\mu$m-thick gold film evaporated on a sapphire substrate. The spherical mirror had a diameter of 35.3~mm and a curvature of 30~mm. The flat mirror consisted of three concentric electrodes forming the Corbino disk with radia 7, 9.9, and 12.9~mm and 5~$\mu$m-wide gap between electrodes. The distance between the two mirrors was $D=13$~mm and determined the frequency of the resonant TEM$_{002}$ mode used in this experiment ($\omega_r/2\pi\approx 35$~GHz). Here we use standard notation for TEM$_{00m}$ mode, according to which $m=0$ corresponds to the fundamental mode of a Fabry-Perot resonator~\cite{kogelnik1966laser}. To excite this mode, the linearly-polarized microwave radiation was supplied from a room temperature source and transmitted into the cell through a fundamental-mode (WR-28) rectangular waveguide, which was vacuum-sealed with a Kapton film K, see Fig.~\ref{fig:1}(a). In addition, the waveguide had an infra-red filter F installed at the 4 K stage of the dilution refrigerator in order to stop thermal radiation from the room temperature. The MW radiation was coupled from the waveguide into the cell through a Kapton-sealed 1.8~mm round aperture made in the middle of the spherical mirror. The coupling was adjusted by changing the thickness of the wall of the mirror where the aperture was made. 

The helium was condensed in the cell such that the liquid level was placed at a distance $h=2.1$~mm above the flat mirror in order to coincide with the position of the first antinode of the MW electric field of the TEM$_{002}$ mode, see Fig.~\ref{fig:1}(c). The liquid level was monitored by observing the downshift of the resonant frequency of the cavity $\omega_r$ as the cell was filled with liquid and comparing this with the shift calculated using a finite element method (FEM). The electrons were produced by thermal emission from a tungsten filament placed above the liquid surface and a 2DES was created and confined on the surface above the flat mirror by applying a positive bias to the central and middle electrodes of the Corbino disk. To excite the cyclotron resonance (CR) of the electrons, a static magnetic field $B$ was applied perpendicular to the liquid helium surface. The value of $B$ was adjusted such that the cyclotron frequency $\omega_c=eB/m_e$, where $e>0$ is the electron charge and $m_e$ is the electron mass, was close to $\omega_r$. In the experiment, both $\omega_c$ and the frequency of the MW radiation $\omega/2\pi$ introduced into the cell could be varied, and either the MW power reflected from the cavity or the dc conductivity response of electrons could be measured as a function of $\omega_c$ and $\omega$. To measure the reflected power we used a pulse-modulated (at frequency $f_m=10$~kHz) MW signal applied to the resonator. The signal reflected from the cavity passed through a cryogenic circulator and was then directed onto a cryogenic InSb detector (QMC Instruments Ltd.) operating at the temperature of the mixing chamber. The detector signal, which was proportional to the incident MW power, was measured by a lock-in amplifier at the modulation frequency $f_m$. The dc conductivity signal of electrons was measured by the standard capacitive (Sommer-Tanner) method using the Corbino disk. To do this, a low-frequency ac signal at 1117~Hz was applied to the inner Corbino electrode. The ac current induced in the middle Corbino electrode by the electron motion was then measured using a lock-in amplifier.    

\section{Results}

\begin{figure}
	\centering
	\includegraphics[width=7cm]{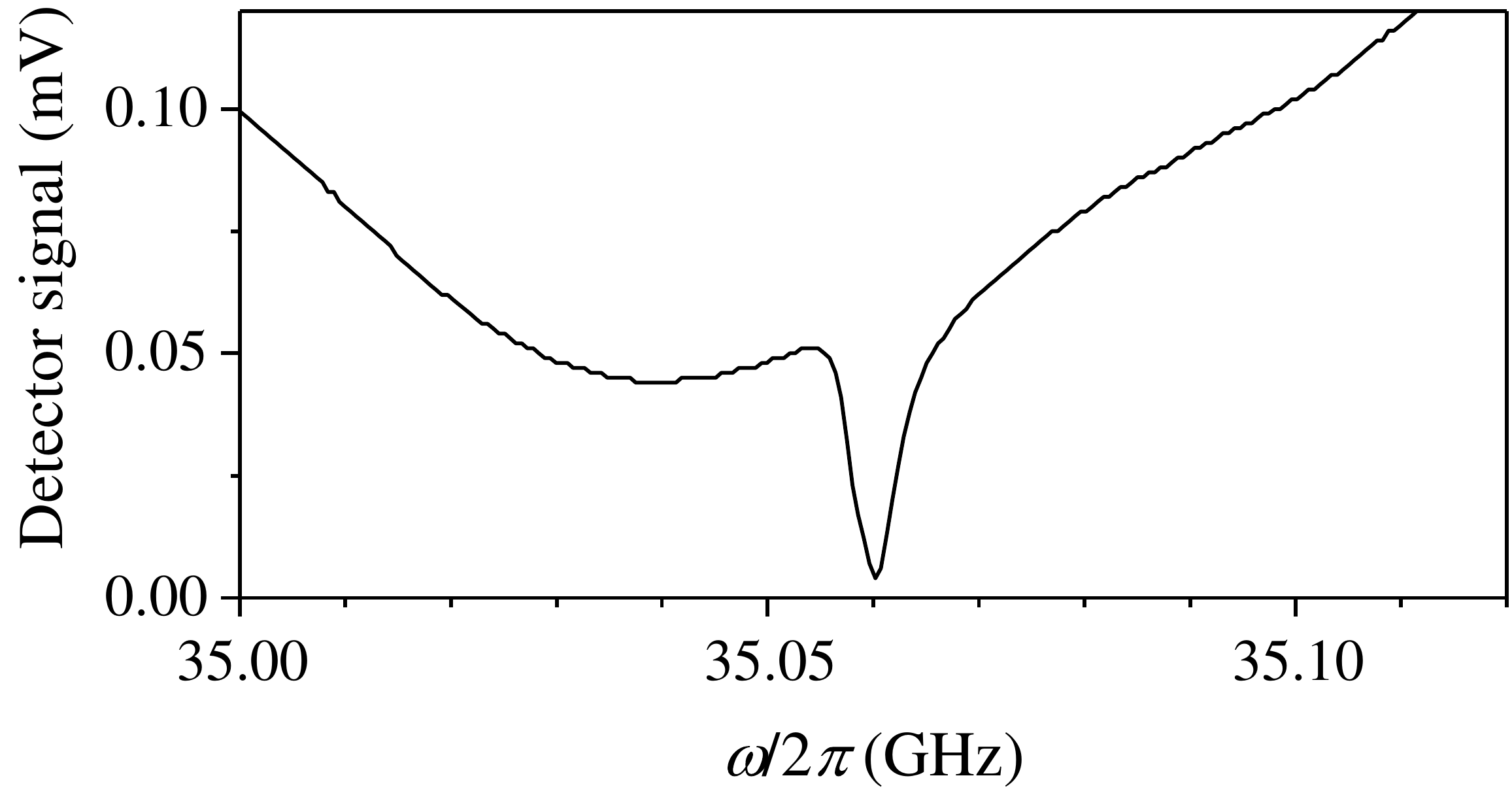}
	\caption{(color online) Spectrum of power reflection from the filled cavity resonator measured at $T=0.2$~K without electrons and an input MW power $P=0$~dBm.}
	\label{fig:2}       
\end{figure}

Figure~\ref{fig:2} shows the reflection spectrum of the filled cavity without a 2DES measured at $T=0.2$~K, $B=0$, and an input MW power of $P=0$~dBm. The spectrum was measured by varying the frequency of the applied MW radiation $\omega$ and recording the InSb detector signal. A sharp dip at $\omega/2\pi\approx 35.06$~GHz is due to the resonant TEM$_{002}$ mode excited in the cavity. Variation of the background with $\omega$ is due to standing wave formation in the transmission line between the cavity and detector as a result of their imperfect matching to the impedance of the transmission line. From the width of the resonance we estimate the quality factor of the cavity to be $Q\approx 9,000$ which is an order-of-magnitude improvement to our previous experiment~\cite{abdurakhimov2016strong}.  

\begin{figure}
\centering
\includegraphics[width=7cm]{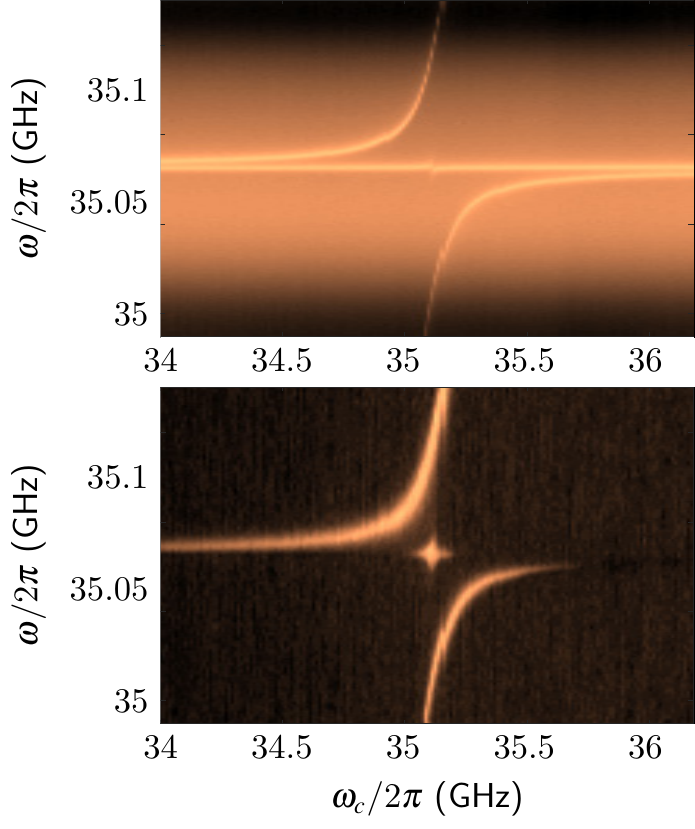}
\caption{(color online) Power reflection from the cavity (top panel) and electron dc conductivity response (bottom panel) versus the cyclotron frequency of electrons $\omega_c$ and frequency of MW excitation $\omega$ measured at $T=0.2$~K for electron surface density $n_s= 8.0\times 10^7$~cm$^{-2}$ and input MW power $P=-9$~dBm.}
\label{fig:3}       
\end{figure}


Figure~\ref{fig:3} (top panel) shows the power reflection from the cavity containing a 2DES with a surface density of $n_s=8.0\times 10^7$~cm$^{-2}$ measured at $T=0.2$~K and an input MW power $P=-9$~dBm. In this experiment, the detector signal was recorded by scanning the frequency of the input MW signal $\omega$ for a fixed value of magnetic field $B$, and therefore a fixed cyclotron frequency $\omega_c$. The experiment was then repeated for different values of $B$. Due to an order-of-magnitude higher cavity quality factor $Q$~\cite{abdurakhimov2016strong}, we can clearly resolve two modes in the reflection spectrum. One mode shows pronounced normal-mode splitting when the cyclotron frequency is close to the resonant frequency of the cavity $\omega_r/2\pi\approx 35.06$~GHz. The other mode shows a single dip when the excitation frequency $\omega$ is close to the resonant frequency $\omega_r$ and is nearly unaffected by the presence of electrons. It is clear that these two modes can be associated with the two circular-polarized components of the input linear-polarized MW signal. For a given direction of the perpendicular magnetic field $B$, only one of the two components (CR-active component) can excite the cyclotron resonance in 2DES, while another component (CR-passive component) can not affect the electron motion in the rotating wave approximation. Thus, the two modes shown in the reflection spectrum can be associated with two circular-polarized components of the MW field in the cavity.     

Figure~\ref{fig:3} (bottom panel) shows the dc conductivity response of a 2DES measured in the same conditions as the power reflection shown in the left panel. In this experiment, an electrical current induced by the electron motion on the middle electrode of the Corbino disk was measured while a low-frequency driving voltage with an amplitude of 20~mV was applied to the center electrode. Unlike the power reflection measurements, which probes the coupled motion of the MW field in the cavity, in this experiment we probe the coupled motion of the electron system. Such motion is strongly affected by only the CR-active component of the MW field. The scattering of electrons during their cyclotron motion introduces heating of the 2DES. Such heating strongly affects the dc conductivity of electrons, which causes a change in the electron current detected by the Corbino disk. Correspondingly, a strong dc conductivity response of the 2DES is observed at the same $\omega_c$ and $\omega$ as the power reflection spectrum of the CR-active mode, c.f., two panels in Fig.~\ref{fig:3}. 

A surprising feature is the appearance of a resonant response of the 2DES at $\omega_c\approx \omega \approx \omega_r$. Such a resonance is also observed in the reflection spectrum of the CR-passive mode, see top panel. We will discuss this unexpected feature in Section~\ref{additional} of this paper.

\section{Classical model}\label{sec:classical}

\begin{figure}
	\begin{center}
		\includegraphics[width=8.5cm]{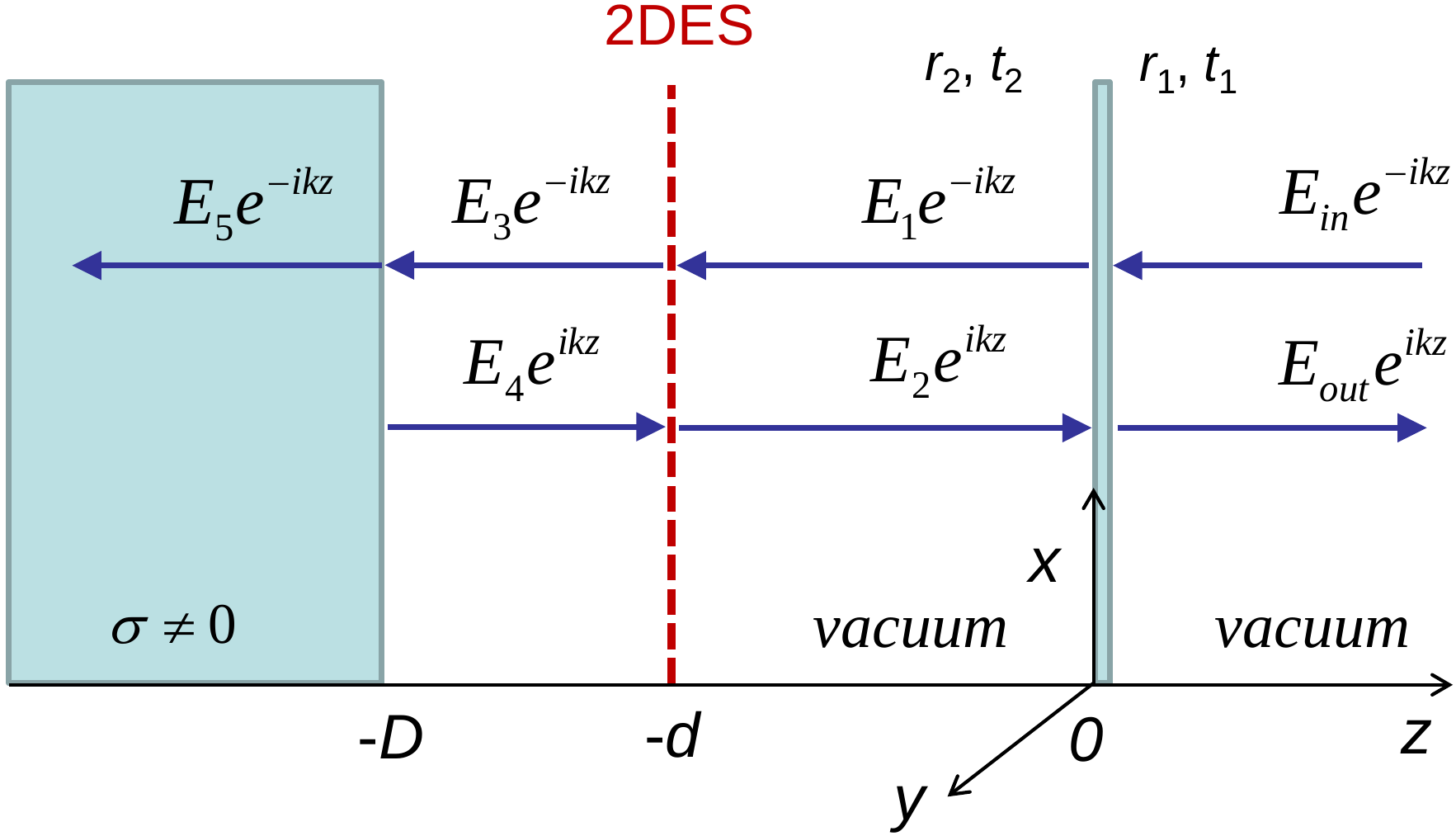}
	\end{center}
	\caption{(color online) Simplified model of the Fabry-Perot resonator containing a 2DES as described in the text. Thick arrows (blue) indicate the direction of propagation for the different components of the EM field inside and outside the resonator excited by the incoming field $\textbf{E}_\textrm{in}e^{-i(kz+\omega t)}$.}
	\label{fig:4}       
\end{figure}

In order to account for the observed coupled electron-field motion we use a model of a 2DES in a simplified Fabry-Perot resonator formed by two infinitely large mirrors located at a distance $D$ apart, see Fig.~\ref{fig:4}. Our treatment is similar to that reported previously~\cite{shikin2002,abdurakhimov2016strong} but properly accounts for two independent polarization degrees of freedom of the cavity field, which is crucial for correct interpretation of our experimental results. The mirrors located at $z=0$ is partially-reflecting with reflection coefficients $r_1$ and $r_2$ for MWs incident on the mirror from $z>0$ and $z<0$, respectively. The corresponding transmission coefficients are $t_1=1+r_1$ and $t_2=1+r_2$. The second mirror (occupying the half-space at $z<-D$) is a good conductor with a finite electrical conductivity $\sigma$ that accounts for internal (Ohmic) losses of the MW field within the cavity. An infinitely large 2DES is located at $z=-d$, $d<D$, and oriented parallel to the plane of the mirrors, see Fig.~\ref{fig:3}. 

We will follow the standard convention and represent the components of our electro-magnetic (EM) fields by complex functions with time dependence in the form $e^{-i\omega t}$. As usual, the real-valued physical quantities measured in an experiment are given by the real part of the corresponding complex-valued expressions, as discussed later in this section. In our model, an input (plane wave) MW radiation propagating in $z$-direction and described by the vector of electric field $\textbf{E}_\textrm{in}$ is incident on the partially-reflecting mirror from $z>0$ and is partially transmitted into the resonator. In order to account for components of the MW field corresponding to two independent circular polarizations, it is convenient to introduce the standard notation $E^{\pm}=(E_x\pm iE_y)/\sqrt{2}$, where $E_x$ and $E_y$ are two components of the complex amplitude (phasor) of the electric field. The complex amplitude $E^+$ ($E^-$) corresponds to the electric field rotating counter-clockwise (clockwise) in the $xy$-plane when looking along the positive $z$-axis. Correspondingly, we will call fields with amplitudes $E^+$ and $E^-$ as left-handed circularly polarized (LHCP) and right-handed circularly polarized (RHCP), respectively.     

The EM field distribution inside and outside of the resonator can be solved classically by considering the superposition of propagating waves and accounting for the boundary conditions at $z=-d$ and $-D$. Designating left- and right-propagating fields inside and outside resonator as indicated in Fig.~\ref{fig:4}, we can write

\begin{subequations}
\begin{align}
& E_1 = t_1 E_{\mathrm{in}} + r_2 E_2, \label{eq:1-1} \\
& E_{\mathrm{out}}  =r_1 E_{\mathrm{in}}  + t_2 E_2 , \\
& E_1  e^{ikd} + E_2  e^{-ikd} = E_3  e^{ikd} + E_4  e^{-ikd},  \\
& -E_3  e^{ikd} + E_4  e^{-ikd} + E_1  e^{ikd} - E_2  e^{-ikd}  = \eta_0 j_{\pm},  \\
& E_3  e^{ikD} + E_4  e^{-ikD} = E_5  e^{i\kappa D}, \\
& -E_3  e^{ikD} + E_4  e^{-ikD} = -\frac{\eta_0}{\eta} E_5  e^{i\kappa D},
\end{align} \label{eq:sys} 
\end{subequations}

\noindent where $E\equiv E^\pm$ is used for shorter notation. Here, we introduce the notation $j_{\pm}=2^{-1/2}(j_x \pm ij_y)$, where $j_x$ and $j_y$ are complex amplitudes of the current density in the 2DES induced by the MW electric field (as mentioned earlier, we assume time dependence in the form $e^{-i\omega t}$), $\eta_0=\sqrt{\mu_0/\varepsilon_0}=377$~Ohm is the intrinsic impedance of vacuum, $k=\omega/c=\omega\sqrt{\varepsilon_0\mu_0}$ is the vacuum propagation constant, $\kappa = \sqrt{\mu_0}\omega/\eta$ is the propagation constant within the conductor, and $\eta$ is the intrinsic impedance of the conductor:

\begin{equation}
\eta \approx \sqrt{\frac{\omega\mu_0}{2\sigma}} (1-i), \quad \frac{1}{\eta_0}\sqrt{\frac{\omega\mu_0}{2\sigma}}<<1.
\label{eq:aproxeta}
\end{equation}    

\noindent The third and forth lines in Eq.(\ref{eq:sys}) express the continuity of electric field and the discontinuity of magnetic field, respectively, at $z=-d$. The latter is due to non-zero electric surface current in the 2DES. The fifth and sixth lines express the continuity of electric and magnetic fields, respectively, at $z=-D$. For the sake of simplicity, we assume that the dielectric constant of liquid helium is equal to 1.    

\indent From Eq.~(\ref{eq:sys}) we can obtain relations between the E-field in the cavity at $z=-d$ for each of the two circular-polarized modes, $E^{\pm}=E_1^{\pm} e^{ikd} + E_2^{\pm} e^{-ikd}$, and the corresponding components  $j_{\pm}$ of the electron current density. Arithmetic is significantly simplified  if we consider the frequency $\omega$ to be close to $\omega_0=c\pi (m+1)/D$, where $m=0,1,2,..$ is the cavity mode number. Note that for an empty cavity each mode is twice degenerate with respect to two independent polarization modes $E^+$ and $E^-$. In addition, we consider that the 2DES to be located at a distance $\lambda_0/4=c\pi/(2\omega_0)$ from the second mirror, that is at the antinode of the electric field. Finally, we assume that $r_1\approx 1$ (that is $t_1\approx 2$) and $r_2\approx -1$ (that is $t_2<<1$). Expanding to first order of $(\omega-\omega_0)/\omega_0$, $\sqrt{\omega\mu_0/(2\sigma)}/\eta_0$, and $t_2$, it is straightforward to obtain the required relation

\begin{equation}
\frac{D}{c} \Big( i(\omega - \omega_r) - (\gamma_{int}+\gamma_{ext})\Big) E^{\pm} - \eta_0 j_{\pm} = 2i(-1)^{(m+1)} E_{in}^{\pm},
\label{eq:first}
\end{equation}

\noindent where $\omega_r=\omega_0 - \delta\omega_{int} - \delta\omega_{ext}$ is the resonant frequency of the cavity, and

\begin{equation}
\delta\omega_{int} = \frac{\omega_0}{\pi (m+1)}\sqrt{\frac{\omega\varepsilon_0}{2\sigma}}, \quad \delta\omega_{ext} = - \textrm{Im}\left( \frac{\omega_0}{2\pi (m+1)} t_2 \right),
\label{eq:shifts}
\end{equation}

\noindent and

\begin{equation}
\quad \gamma_{int} = \frac{\omega_0}{\pi (m+1)}\sqrt{\frac{\omega\varepsilon_0}{2\sigma}}, \quad \gamma_{ext}=\textrm{Re}\left( \frac{\omega_0}{2\pi (m+1)} t_2 \right),
\label{eq:gammas}
\end{equation}

\noindent are the internal (Ohmic) and external (radiative) loss rates of the resonator, respectively.

The second relation between $E^{\pm}$ and $j_{\pm}$ is given by the definition of ac conductivity, $j_{\pm}=\sigma_{\pm} E^{\pm}$. The expression for $\sigma_{\pm}$ can be easily obtained from the classical equations of motion for a collection of point-charge particles having the surface density $n_s$, by taking into account the Lorenz force due to the perpendicular magnetic field $\textbf{B}$ and ignoring the Coulomb interaction between particles. For certainty, we assume that the applied magnetic field is in the positive $z$-direction. From

\begin{equation}
m_e\frac{d \textbf{v}}{dt} = -e\textbf{E} - e\textbf{v} \times \textbf{B} - m_e \textbf{v} \nu,
\label{eq:Newt}
\end{equation}  

\noindent where $\textbf{v}$ is the electron velocity parallel to the liquid helium surface and $\nu$ is the (phenomenological) scattering rate of electrons, we can write the equation of motion for the electron current density $\textbf{j}=-en_s\textbf{v}$ and obtain the following expression for the ac conductivity

\begin{equation}
\sigma_{\pm}=\frac{n_se^2}{m_e}\frac{1}{\nu-i(\omega\pm \omega_c)}.
\label{eq:sigm}
\end{equation} 

\noindent Note that the same form of expression can be obtained using a quantum treatment and taking into account the Coulomb interaction between electrons~\cite{monarkha2013two}. 

From Eqs. (\ref{eq:first}), (\ref{eq:sigm}) and the definition $j_{\pm}=\sigma_{\pm} E^{\pm}$  we obtain a system of coupled equations for electron-field motion

\begin{equation}\begin{split}
&\begin{pmatrix}
\frac{D}{c} \Big[ i(\omega - \omega_r) - (\gamma_{int}+\gamma_{ext})\Big] \hfill &  \hfill -\eta_0 \\
n_se^2/m_e & \hfill i(\omega\pm \omega_c)-\nu \hfill
\end{pmatrix}
\begin{pmatrix}
E^{\pm} \\
j_{\pm} 
\end{pmatrix}\\
& =
\begin{pmatrix}
2i(-1)^{m+1} E_{in}^{\pm} \\
0 
\end{pmatrix}
\end{split}\label{eq:classical-coupling}\end{equation}

\noindent In the absence of external drive, $E_{in}^{\pm}=0$, the nontrivial solutions for $E^{\pm}$ and $j_{\pm}$ only exist for $\omega$ that cause the determinant of the left-hand-side matrix of Eq.~(\ref{eq:classical-coupling}) to vanish. This provides us with frequencies $\omega_{1,2}$ for the normal (eigen) modes of the coupled electron-field motion. It is instructive to find these frequencies for the case of zero losses, that is $\nu=0$ and $\gamma_{int}+\gamma_{ext}=0$. Then, we obtain

\begin{equation}
(\omega - \omega_r)(\omega \pm \omega_c) - \frac{n_se^2}{m_e\varepsilon_0 D} = 0.
\end{equation}

\noindent For $\omega_c\approx \omega_r$, two solutions $\omega_{1,2}=\omega_r \pm g$, where 

\begin{equation}
g=\sqrt{\frac{n_s e^2}{m_e\varepsilon_0 D}},
\label{eq:g}
\end{equation}

\noindent are realized for the $E^-$ component of the microwave field. For this component, the normal-mode splitting in the spectrum of coupled electron-field motion is given by $2g$. $E^-$ corresponds to a RHCP electric field that rotates in the same direction as an electron in the static $B$-field oriented in the positive $z$-direction. In other words, the co-rotating component $E^-$ corresponds to the CR-active component of cavity mode. The counter-rotating (CR-passive) component, $E^+$, does not show any splitting as expected.   

\indent For the sake of comparison with our experimental results we derive an expression for the normalized power reflection, which we define as the ratio between the time-averaged output and input MW powers, $P_\textrm{R}=\textbf{E}_{out}\textbf{E}_{out}^*/(\textbf{E}_{in}\textbf{E}_{in}^*)$. From Eq.~(\ref{eq:sys}) we obtain

\begin{equation}
E_{out}^{\pm}=\left( 1 - \frac{ 2(\delta\omega_{ext} + i\gamma_{ext}) }{ (\omega -\omega_r) + i\gamma  + i\eta_0\sigma_{\pm} c/D } \right) E_{in}^{\pm},
\label{eq:outCORR}
\end{equation}

\noindent where $\gamma=\gamma_{int}+\gamma_{ext}$ is the total loss rate of the cavity. Assuming a linearly polarized (along $x$-axis) input MW field with $E_{in}^+=E_{in}^-=E_0/\sqrt{2}$, we obtain
\begin{equation} \begin{split}
	P_\textrm{R}=&\frac{\left| E_{out}^+\right|^2 + \left| E_{out}^-\right|^2}{E_0^2} 
	\\
	=&\frac{1}{2} \left| 1 + \frac{ 2(\gamma_{ext} - i\delta\omega_{ext}) }{ i(\omega -\omega_r) - \gamma - \frac{\sigma_{+}}{\varepsilon_0 D}} \right|^2 \\ 
	&+ 
	\frac{1}{2}\left| 1+ \frac{ 2(\gamma_{ext} - i\delta\omega_{ext}) }{ i(\omega -\omega_r) - \gamma - \frac{\sigma_{-}}{\varepsilon_0 D}}  \right|^2 . 
	\label{eq:powR}
\end{split} \end{equation}

\noindent Similarly, the time-averaged power of Joule heating in the 2DES due to the MW electric field is given by

\begin{equation}\begin{split}
P_\textrm{J}=&\langle \textrm{Re}(\textbf{j})\textrm{Re}(\textbf{E}) \rangle_t \\
=&\frac{1}{2} \left( \textrm{Re}(\sigma_+)|E^+|^2 + \textrm{Re}(\sigma_-)|E^-|^2 \right).
\label{eq:powJ}
\end{split}\end{equation}

\begin{figure}
\begin{center}
\includegraphics[width=7cm,keepaspectratio]{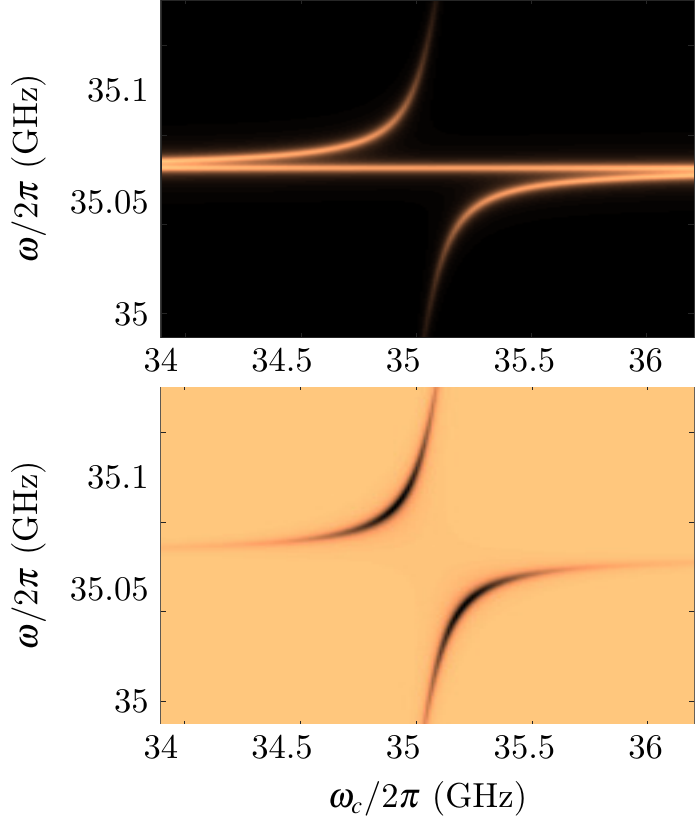}
\end{center}
\caption{(color online) ower reflection from the cavity (top panel) and power of the Joule heating of the 2DES by the MW field (bottom panel) versus the cyclotron frequency of electrons $\omega_c$, and the frequency of MW excitation $\omega$ calculated from Eqs.~(\ref{eq:classical-coupling}), (\ref{eq:powR}) and (\ref{eq:powJ}) for $n_s=6\times 10^7$~cm$^{-2}$, $\nu=8\times 10^7$~s$^{-1}$ and $Q=20,000$.}
\label{fig:5}       
\end{figure}

The numerical solutions for $P_\textrm{R}$ and $P_\textrm{J}$ obtained by solving Eq.~(\ref{eq:classical-coupling}) for $n_s=6\times 10^7$~cm$^{-2}$, $\nu=8\times 10^7$~s$^{-1}$ and $Q=20,000$ are shown in Fig.~\ref{fig:5}. Comparing these solutions to Fig.~\ref{fig:3}, we can see that our completely classical model reproduces the main features of the experimental results. In particular, it reproduces the normal-mode splitting observed in both the cavity field and electron system responses. 

\section{Full quantum treatment}

As follows from the model described above, the normal-mode spitting is given by a completely classical expression, Eq.~(\ref{eq:g}), that does not contain $\hbar$. On the other hand, our expression gives the correct $\sqrt{N}$ enhancement of the coupling between the cavity field and an $N$-particle ensemble.~\cite{hagenmuller2010ultrastrong,scalari2012ultrastrong,zhang2016collective,li2018bloch} It is easy to see that we can bring our classical expression to a QED form expressed in terms of the rms electric field of vacuum, $E_\textrm{vac}=\sqrt{\hbar\omega/2\varepsilon_0 V}$, by employing a simple trick of multiplying and dividing Eq.~(\ref{eq:g}) by $\hbar$. Indeed, in this case we obtain

\begin{equation}
g=\frac{e}{\hbar}\sqrt{\frac{2\hbar}{m_e \omega_c}}\sqrt{\frac{\hbar\omega_c}{2\varepsilon_0 V}}\sqrt{n_s S}= \frac{\sqrt{2}el_BE_\textrm{vac}}{\hbar}\sqrt{N},
\label{eq:trick}
\end{equation}

\noindent where $l_B=\sqrt{\hbar/eB}$ is the magnetic length. Below we show that this result can be reproduced by the full quantum-mechanical model.

\subsection{Operator description}

Our general approach is similar to that described in the literature~\cite{hagenmuller2010ultrastrong,julsgaard2012dynamical,julsgaard2013quantum,li2018bloch}. We start with the description of an EM field inside the cavity shown in Fig.~\ref{fig:4}. The field inside an empty single-mode cavity can be described by an operator of vector potential 
\begin{equation}
\hat{\*A}(z,t)= 
\sqrt{\frac{\hbar\omega_0}{2\epsilon_0 V}} 
\sum_{\alpha}
\left( \*e_{\alpha} f(z) \hat{a}_{\alpha} + \*e_{\alpha}^*  f^*(z) \hat{a}_{\alpha}^\dagger \right),
\end{equation}

\noindent where the sum is over two polarization degrees of freedom described by unit vectors $\*e_\alpha$. For example, $\*e_{\pm}=(2^{-1/2})(\*e_x \mp i\*e_y)$, where $\*e_x$ and $\*e_y$ are the unit vectors in the $x$ and $y$-direction, respectively, represent the LHCP and RHCP fields defined in the previous section. We will use notations $\hat{a}_\textrm{L}$ and $\hat{a}_\textrm{R}$ for the corresponding photon operators. The normalized function $f(z)=i\sqrt{2}\sin(k_0 z)$, where $k_0=\pi(m+1)/D$, $m=0,1,2,..$ , describes the field distribution of given mode~\cite{meystre2007quantum}.


For an ensemble of free 2D electrons in a perpendicular static magnetic field $\*B_0=B\*e_z$, it is convenient to introduce the operator of kinematic momentum $\pmb{\hat{\pi}}=\hat{\textbf{p}} + e\hat{\*A}_0$ for a single electron, where $\hat{\*A}_0=(-\hat{y} B/2, \hat{x} B/2,0)$ is the vector potential of $\*B_0$. The commutation relation $[\hat{\pi}_x,\hat{\pi}_y]=-i\hbar e B $ leads to the definition of a dimensionless annihilation operator 

\begin{equation} 
\hat{b}=\sqrt{\frac{1}{2\hbar  e B}}\qty(\hat{\pi}_x-i\hat{\pi}_y), 
\end{equation}

\noindent that satisfies the commutation relation $[\hat{b},\hat{b}^\dagger]$=1. This single-particle operator can be related to a complex current density operator for an $N$-particle system $\hat{j}_-=2^{-1/2}(\hat{j}_x - i\hat{j}_y)$. Here $ \hat{j}_{x(y)} = (-e/m_e S) \sum\limits_e \hat{\pi}_{x(y)}$, where the sum is over all electrons in the system, and $S$ is the surface area occupied by the system. The operators $\hat{b}$ and $\hat{j}_-$ are related by $\hat{j}_-=-(e\omega_c l_B N/S) \hat{b}$.
 
\subsection{Coupled oscillator model}

Under the Coulomb gauge condition, $\nabla \*A=0$ and $\grad \phi=0$, the Hamiltonian of the system composed of a single EM mode and and $N$-electron system can be written as 

\begin{widetext}
\begin{equation} 
\hat{H}=\hbar\omega_r \sum_{\alpha} \hat{a}_{\alpha}^{\dagger}\hat{a}_{\alpha} + \frac{1}{2m_e} \sum_e \left( \hat{\pmb\pi} + e \hat{\*A} \right)^2\approx \hbar \omega_r \left( \hat{a}_\textrm{L}^\dagger \hat{a}_\textrm{L}  + \hat{a}_\textrm{R}^\dagger \hat{a}_\textrm{R} \right) + \hbar \omega_c \sum_e \hat{b}^\dagger \hat{b} + \hbar g_0 \sum_e \left( \hat{b} \hat{a}_\textrm{R}^\dagger +\hat{b}^\dagger \hat{a}_\textrm{R} \right),
\label{eq:hamil}
\end{equation}
\end{widetext}
 
\noindent where we have adopted the notations used in the previous section for the frequency of resonant cavity mode $\omega_r\approx k_0/c$, and neglected the $\mathbf{A}^2$ term under the rotating wave approximation (RWA). The single-electron coupling constant is given by $g_0 = \sqrt{e^2\omega_c/(m_e\epsilon_0 \omega_r V)}$. The interaction term in the above equation can be viewed as an exchange of a quantum of excitation between the electron cyclotron and the cavity RHCP field. In the RWA, the counter-rotating LHCP field of the cavity mode does not contribute to the interaction. In Section~\ref{additional}, we will reexamine the contribution of the LHCP pumping field to the resonance of electrons in a real resonator. Finally, as in the previous section we assume that the electrons are located in the antinode of the electric field of the EM mode, thus $|f(z_e)|^2=2$.   

Next, we write the Heisenberg equations of motion for the time-dependent operators $\hat{a}_\textrm{L}$ and $\hat{b}$ as 

\begin{subequations}\begin{align}
 \dot{\hat{a}}_\textrm{R} & = (-i\omega_r - \gamma) \hat{a}_\textrm{R} - i g_0 N \hat{b} + \hat{F}_a, \\
\dot{\hat{b}} &= -i g_0 \hat{a}_\textrm{R} + (-i\omega_c - \nu) \hat{b} + \hat{F}_b.  
\end{align} 
\label{eq:heisen}
\end{subequations}    


\noindent Here, we use the quantum Langevin equation and introduce the Langevin noise operators $\hat{F}_a$ and $\hat{F}_b$, which vanish in the corresponding mean value equations, as well as the relaxation rates $\gamma$ and $\nu$, in order to account for the interaction of the system with the environment~\cite{walls2008quantum}. The above equations describe two coupled harmonic oscillators with frequencies $\omega_r$ and $\omega_c$. It is easy to check that the corresponding equations for the mean values of operators $\hat{a}_\textrm{L}$ and $\hat{b}$ obtained from (\ref{eq:heisen}) are equivalent to our classical equations~(\ref{eq:classical-coupling}) for complex amplitudes $E^-$ and $j_-$. The operators corresponding to these quantities are given by the Fourier components of operators $\hat{j}_-$ and $\hat{E}^-=iE_\textrm{vac} \hat{a}_\textrm{R}$. Using equations of motion (\ref{eq:heisen}), we obtain 

\begin{equation} \begin{split}
 \left[i(\omega - \omega_r) -\gamma\right] \langle \hat{E}^-(\omega)\rangle - \frac{E_\textrm{vac}^2 S}{\hbar\omega_r} \langle \hat{j}_-(\omega) \rangle = 0, \\
  \frac{e^2 \omega_c N}{m_e \omega_r S} \langle \hat{E}^-(\omega) \rangle + \left[i(\omega - \omega_c) - \nu\right] \langle \hat{j}_-(\omega) \rangle =0.
\end{split} \end{equation}

\noindent For $\omega_c\approx \omega_r$, the corresponding equations for the mean values of quantum-mechanical operators give the same results as the classical equations (\ref{eq:classical-coupling}), however without the external pumping term. Note that the expression for the eigen mode splitting coincides with Eq.~(\ref{eq:trick}), as expected.  



In order to include external pumping to our model, it is convenient to use Collett and Gardiner's approach, which allows us to obtain a relation between the input and output fields~\cite{collett1984squeezing}. We consider a one sided cavity for which the main source of loss (with loss rate $\gamma$) is the coupling to an external field. In this case, the boundary condition at the coupling port reads   

\begin{equation}
\sqrt{2\gamma} \hat{a}_\textrm{R(L)} (t) = \hat{a}_\textrm{R(L)}^{\mathrm{(in)}}(t)+\hat{a}_\textrm{R(L)}^{\mathrm{(out)}}(t),
\label{eq:inport}
\end{equation}

\noindent which is consistent with boundary conditions \eqref{eq:1-1}. Note that operators for external (in and out) fields are normalized such that $\hat{a}^\dagger \hat{a}$ gives the in(out)coming number of photons per second. The equations of motion for operators $\hat{a}_\textrm{R(L)}$ and $\hat{b}$ lead to the linear algebraic equations for the corresponding Fourier transforms $\*u = (\hat{a}_\textrm{R}(\omega), \hat{a}_\textrm{L}(\omega), \hat{b}(\omega))$, which can be written in matrix form as $M \*u = -\sqrt{2\gamma}\*u^\mathrm{(in)}$, where

\begin{equation}
    M = \begin{pmatrix}
    i(\omega-\omega_r) - \gamma & 0 &  -ig_0N  \\ 
    0&  i(\omega-\omega_r) - \gamma & 0   \\ 
    -ig_0 & 0 & i(\omega-\omega_c) - \nu   \\
\end{pmatrix} .
\end{equation}

\noindent The solution for $\*u^\mathrm{(in)}=(\hat{a}^\mathrm{(in)}_\textrm{R}(\omega), \hat{a}^\mathrm{(in)}_\textrm{L}(\omega), 0)$ can be obtained by simply inverting the matrix $M$, giving us

\begin{subequations}\label{eq:22}\begin{align}
    \hat{a}_\textrm{R}(\omega) 
        &= \frac{\sqrt{2\gamma}(\nu -i (\omega - \omega_c))}
        {\left(i (\omega-\omega_r) - \gamma \right) \left(i (\omega-\omega_c) -\nu \right) + g_0^2 N} \hat{a}_\textrm{R}^\mathrm{(in)}(\omega),\\
    \hat{a}_\textrm{L}(\omega)
        &= \frac{\sqrt{2\gamma}}{\gamma - i (\omega-\omega_r)}\hat{a}_\textrm{L}^\mathrm{(in)}(\omega),\\
    \hat{b}(\omega) 
        &= -\frac{ig_0\sqrt{2\gamma}}
				{\left(i (\omega-\omega_r) -\gamma \right) \left(i (\omega-\omega_c) - \nu \right) + g_0^2 N} \hat{a}_\textrm{R}^\mathrm{(in)}(\omega).
\end{align}\end{subequations}

\noindent Using the above equations together with the boundary condition (\ref{eq:inport}), we obtain the linear input-output relations for two polarization modes

\begin{subequations}\begin{align}
    \frac{\hat{a}_\textrm{R}^\mathrm{(out)}}{\hat{a}_\textrm{R}^\mathrm{(in)}} &= -1 - \frac{2\gamma(i (\omega-\omega_c)-\nu)}
        {\left(i (\omega-\omega_r) -\gamma \right) \left(i (\omega-\omega_c) -\nu \right) + g_0^2 N},\\
    \frac{\hat{a}_\textrm{L}^\mathrm{(out)}}{\hat{a}_\textrm{L}^\mathrm{(in)}} &= \frac{(\omega-\omega_r) - i\gamma}{ (\omega-\omega_r) + i\gamma}. 
\end{align}\end{subequations}

\noindent The normalized power reflection is given by $P_\textrm{R}=\langle \hat{a}^\mathrm{(out)}\dagger \hat{a}^\mathrm{(out)} \rangle / \langle \hat{a}^\mathrm{(in)}\dagger \hat{a}^\mathrm{(in)} \rangle$, which results in the same relations as for classical quantities, see Eq.~(\ref{eq:outCORR}). The time-averaged power absorbed by the electron system from the MW field is given by 

\begin{equation}
P_\textrm{J}=\langle \hat{\textbf{j}}\hat{\textbf{E}} \rangle = -\frac{i\hbar g_0\omega_c n_s}{D} \langle \hat{b}^\dagger \hat{a}_\textrm{R} - \hat{b}^\dagger \hat{a}_\textrm{R}^\dagger \rangle.
\end{equation}

\noindent The above equations completely reproduce the results shown in Fig.~\ref{fig:5}.

\section{Coupling to LHCP pumping field}\label{additional}

We have shown that both classical and full quantum models reproduce the result of the normal mode splitting due to coupling between the cyclotron motion of electrons and co-rotating (RHCP) component of the cavity field, see Fig.~\ref{fig:5}. However, the experimental data shown in Fig.~\ref{fig:3} exhibit an additional resonance peak when the MW frequency $\omega$ is close to both the cavity frequency $\omega_r$ and cyclotron frequency $\omega_c$. This result appears to indicate that there is an effective coupling between the electron cyclotron motion and CR-passive (LHCP) component of the cavity field. Since under the conditions of our experiment the ratio of the collective coupling constant to the cyclotron frequency, $g/\omega_c\sim 10^{-3}$, is small, the rotating wave approximation used in Eq.~(\ref{eq:hamil}) seems to be well justified. The resonant coupling of electrons to the LHCP component of MW field is possible due to the second-order processes accompanied by simultaneous scattering of electrons from ripplons, which, for example, give rise to the observed conductivity response of electrons on helium at the harmonics of the cyclotron resonance.~\cite{zador2018polar} However, this contributes only a small fraction of $\nu/\omega_c\lesssim 10^{-3}$ to the electron conductivity, an effect comparable to the counter-rotating terms neglected in (\ref{eq:hamil}) under the RWA. Thus, it is unlikely that the counter-rotating component of field can cause effects shown in Fig.~\ref{fig:3}. This calls for a detailed examination of the structure of the resonant MW field in the cavity under pumping with the LHCP field. 

\begin{figure}
	\subfloat[\label{fig:6a}]{%
		\includegraphics[width=.48\linewidth]{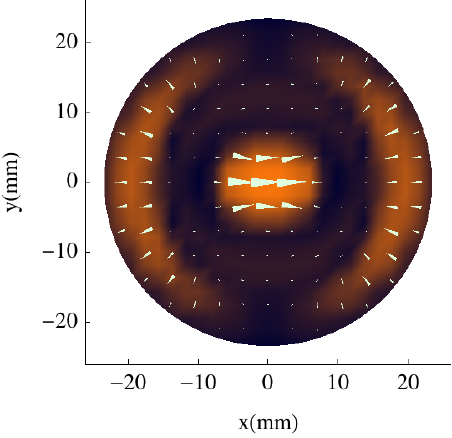}%
	}~
	\subfloat[\label{fig:6b}]{%
		\includegraphics[width=.48\linewidth]{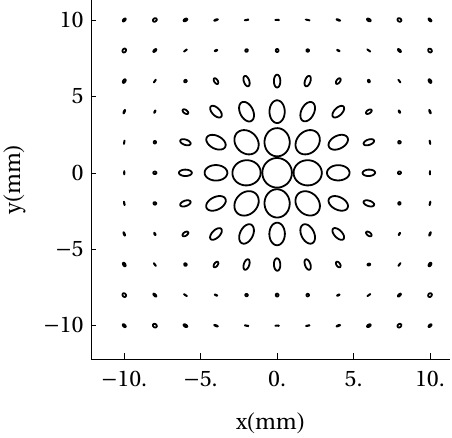}%
	}
	\caption{ (Color online) a) Distribution of microwave electric field on the surface of liquid helium under excitation with linearly ($x$-direction) polarized field. b) Polarization of microwave electric field on the surface of liquid helium under excitation with circular polarized field.}
	\label{fig:6}
\end{figure}

We start by noting that the transverse (TEM) resonant mode shown in Fig.~\ref{fig:3} fails to give an adequate description of the resonant electromagnetic field in our real resonator. Indeed, the confinement of our Fabry-Perot resonator inside the closed cylindrical cell imposes boundary conditions on the microwave electric field inside the cell and makes it impossible to preserve the circular polarization of the pumping field. This is readily seen from the fact that the time-dependent vector of the electric field has to remain perpendicular to the conductive walls, thus it corresponds to the linearly polarized field. As an illustration, Fig.~\ref{fig:6a} shows the distribution of microwave electric field (white arrows) at the surface of liquid helium in our cell when the cavity is pumped by the linearly ($x$-direction) polarized field. Assuming rotation invariance of the calculated field in our axially-symmetric cell, from this figure we can reconstruct the spatial structure of the microwave electric field on the liquid helium surface when the cavity is pumped with a circularly polarized field. A schematic plot is shown in Fig.~\ref{fig:6b}, where each closed contour traces the vector of ac electric field over one cycle of oscillation, while the size of each contour scales with the amplitude of electric field. From this plot it is clear that our cavity preserves the circular polarization of the pumping field only close to the center of the cavity, while the field is elliptically polarized away from the center. This means the cyclotron motion of electrons located away from the center can also have an effective coupling to the mode excited in our cavity by the LHCP field.

\begin{figure}
\centering
\includegraphics[width=7cm]{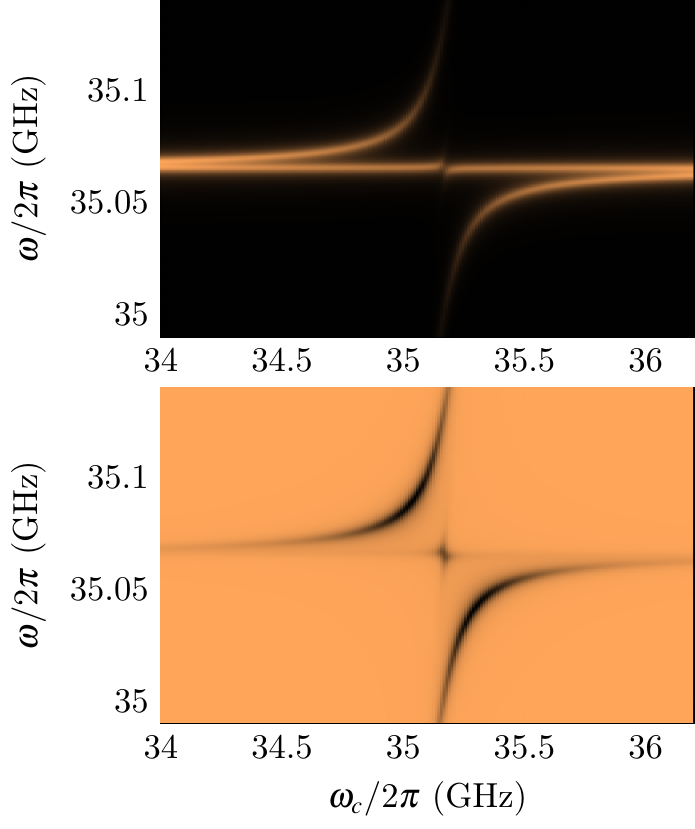}
\caption{(color online) Power reflection from the cavity (top panel) and power of the Joule heating of 2DES by MW field (bottom panel) versus the cyclotron frequency of electrons $\omega_c$ and frequency of MW excitation $\omega$ calculated using model with interaction term described by Eq.~\eqref{eq:26}.}
\label{fig:7}       
\end{figure}


In order to quantitatively account for this effect, we modify the Hamiltonian \eqref{eq:hamil} to reflect the spatial dependence of the coupling constant in the interaction term. The third term in the Hamiltonian \eqref{eq:hamil} should read

\begin{equation}\label{eq:25}
	\hat{H}_I = \sum_i \hbar \left(g_R^{(i)} \hat{b}^{(i)} \hat{a}_R^\dagger +  g_L^{(i)} \hat{b}^{(i)} \hat{a}_L^\dagger \right) + H.c. , 
\end{equation}

\noindent where $g^{(i)}_{R(L)}$ is a function of the position of particle $i$. To simplify numerical calculations, it is convenient to assume a continuous distribution of electron charge on the liquid helium surface and replace summation over particles with integration over the surface. Written in the polar coordinates, the interaction Hamiltonian becomes

\begin{equation}\label{eq:26}
\hat{H}_I = \int_0^{2\pi}  \dd \phi \int_0^{R_0} \,r \dd r\, \hat{\mathbf{A}}(r,\phi)\cdot\hat{\mathbf{j}}(r,\phi), 
\end{equation}

\noindent where $\phi$ is the azimuth, $r$ is the distance from the center of the cell, and $R_0$ is the inner radius of the cell. In our numerical simulations, we assumed a uniform charge distribution within a circle of diameter 10~mm centered at the middle of the cell. The power reflection and the power of the Joule heating were numerically calculated using the mean value equations as described in the previous section, see Eqs.~(\ref{eq:22}). The results are plotted in Fig.~\ref{fig:7}. As one can see, we reproduce the resonance feature at $\omega_c\approx \omega \approx \omega_r$ in both the power reflection spectrum for the LHCP component of pumping field and electron photoconductivity response, c.f. Fig.~\ref{fig:3}.   
  
\section{Conclusions}

We have demonstrated strong coupling between the cyclotron motion of a 2DES on liquid helium and the co-rotating polarization component of the electromagnetic mode of the Fabry-Perot resonator. The effect is manifested by a pronounced splitting in the eigenspectrum of coupled motion, which was observed in both the cavity reflection signal and the electron photoconductivity response. This observation was completely accounted for by the classical equations of motion for electromagnetic fields in a cavity. For the sake of comparison, we have demonstrated complete agreement between the results obtained from both the classical electrodynamic and cavity-QED treatments. The essential physics of the system is completely described by a model of coupled harmonic oscillators. The linearity of the obtained equations of motion for quantum-mechanical operators allows one to construct closed mean-value equations for observables which correspond to the classical equations of motion. This confirms the classical nature of the $\sqrt{N}$ enhancement of the normal mode splitting for a many-particle ensemble. Such a result should not be surprising since the input EM field is in a coherent state and there are no considerable nonlinear effects in our experiment. This result is in accord with the earlier discussions in atomic physics and quantum optics. We note that similar conclusions have been recently reached in classical and semiclassical studies of the strong coupling regime in solid-state systems~\cite{bai2015spin,todorov2012prb}. Still, in many cases the full quantum treatment is simple comparing with rather tedious classical approach. For example, we used the quantum approach in Section~\ref{additional} to successfully reproduce the additional resonance observed in our experiment even though its origin is purely classical.  

For the counter-rotating polarization component of the MW field, we have observed a resonance in both the cavity and electron responses when the MW frequency coincides with both the cavity and cyclotron frequencies. As we have shown, this surprising feature, which can be easily misinterpreted as a signature of weak coupling of electron ensemble to the CR-passive component of EM field, arises because of the mixing between two circular polarization components of the cavity field due to its interaction with conductive walls of the cell. Thus, we demonstrate that understanding the structure of electromagnetic field in a resonator is important for correct interpretation of our and similar experiments dealing with collective coupling between particle ensembles and cavity field, while the quantum model is useful for a succinct formulation of the problem.           

\begin{acknowledgments}
The work was supported by an internal grant from the Okinawa Institute of Science and Technology (OIST) Graduate University. We thank A.~D. Chepelianskii for helpful discussions and V.~P. Dvornichenko for technical support.  
\end{acknowledgments}

\bibliographystyle{apsrev4-1}
\bibliography{strongcoupling2018}

\begin{thebibliography}{50}%
\makeatletter
\providecommand \@ifxundefined [1]{%
 \@ifx{#1\undefined}
}%
\providecommand \@ifnum [1]{%
 \ifnum #1\expandafter \@firstoftwo
 \else \expandafter \@secondoftwo
 \fi
}%
\providecommand \@ifx [1]{%
 \ifx #1\expandafter \@firstoftwo
 \else \expandafter \@secondoftwo
 \fi
}%
\providecommand \natexlab [1]{#1}%
\providecommand \enquote  [1]{``#1''}%
\providecommand \bibnamefont  [1]{#1}%
\providecommand \bibfnamefont [1]{#1}%
\providecommand \citenamefont [1]{#1}%
\providecommand \href@noop [0]{\@secondoftwo}%
\providecommand \href [0]{\begingroup \@sanitize@url \@href}%
\providecommand \@href[1]{\@@startlink{#1}\@@href}%
\providecommand \@@href[1]{\endgroup#1\@@endlink}%
\providecommand \@sanitize@url [0]{\catcode `\\12\catcode `\$12\catcode
  `\&12\catcode `\#12\catcode `\^12\catcode `\_12\catcode `\%12\relax}%
\providecommand \@@startlink[1]{}%
\providecommand \@@endlink[0]{}%
\providecommand \url  [0]{\begingroup\@sanitize@url \@url }%
\providecommand \@url [1]{\endgroup\@href {#1}{\urlprefix }}%
\providecommand \urlprefix  [0]{URL }%
\providecommand \Eprint [0]{\href }%
\providecommand \doibase [0]{http://dx.doi.org/}%
\providecommand \selectlanguage [0]{\@gobble}%
\providecommand \bibinfo  [0]{\@secondoftwo}%
\providecommand \bibfield  [0]{\@secondoftwo}%
\providecommand \translation [1]{[#1]}%
\providecommand \BibitemOpen [0]{}%
\providecommand \bibitemStop [0]{}%
\providecommand \bibitemNoStop [0]{.\EOS\space}%
\providecommand \EOS [0]{\spacefactor3000\relax}%
\providecommand \BibitemShut  [1]{\csname bibitem#1\endcsname}%
\let\auto@bib@innerbib\@empty
\bibitem [{\citenamefont {Dicke}(1954)}]{Dicke1954}%
  \BibitemOpen
  \bibfield  {author} {\bibinfo {author} {\bibfnamefont {R.~H.}\ \bibnamefont
  {Dicke}},\ }\href@noop {} {\bibfield  {journal} {\bibinfo  {journal} {Phys.
  Rev.}\ }\textbf {\bibinfo {volume} {93}},\ \bibinfo {pages} {99} (\bibinfo
  {year} {1954})}\BibitemShut {NoStop}%
\bibitem [{\citenamefont {Kaluzny}\ \emph {et~al.}(1983)\citenamefont
  {Kaluzny}, \citenamefont {Goy}, \citenamefont {Gross}, \citenamefont
  {Raimond},\ and\ \citenamefont {Haroche}}]{kaluzny1983observation}%
  \BibitemOpen
  \bibfield  {author} {\bibinfo {author} {\bibfnamefont {Y.}~\bibnamefont
  {Kaluzny}}, \bibinfo {author} {\bibfnamefont {P.}~\bibnamefont {Goy}},
  \bibinfo {author} {\bibfnamefont {M.}~\bibnamefont {Gross}}, \bibinfo
  {author} {\bibfnamefont {J.~M.}\ \bibnamefont {Raimond}}, \ and\ \bibinfo
  {author} {\bibfnamefont {S.}~\bibnamefont {Haroche}},\ }\href {\doibase
  10.1103/PhysRevLett.51.1175} {\bibfield  {journal} {\bibinfo  {journal}
  {Phys. Rev. Lett.}\ }\textbf {\bibinfo {volume} {51}},\ \bibinfo {pages}
  {1175} (\bibinfo {year} {1983})}\BibitemShut {NoStop}%
\bibitem [{\citenamefont {Raizen}\ \emph {et~al.}(1989)\citenamefont {Raizen},
  \citenamefont {Thompson}, \citenamefont {Brecha}, \citenamefont {Kimble},\
  and\ \citenamefont {Carmichael}}]{raizen1989normal}%
  \BibitemOpen
  \bibfield  {author} {\bibinfo {author} {\bibfnamefont {M.~G.}\ \bibnamefont
  {Raizen}}, \bibinfo {author} {\bibfnamefont {R.~J.}\ \bibnamefont
  {Thompson}}, \bibinfo {author} {\bibfnamefont {R.~J.}\ \bibnamefont
  {Brecha}}, \bibinfo {author} {\bibfnamefont {H.~J.}\ \bibnamefont {Kimble}},
  \ and\ \bibinfo {author} {\bibfnamefont {H.~J.}\ \bibnamefont {Carmichael}},\
  }\href {\doibase 10.1103/PhysRevLett.63.240} {\bibfield  {journal} {\bibinfo
  {journal} {Phys. Rev. Lett.}\ }\textbf {\bibinfo {volume} {63}},\ \bibinfo
  {pages} {240} (\bibinfo {year} {1989})}\BibitemShut {NoStop}%
\bibitem [{\citenamefont {Zhu}\ \emph {et~al.}(1990)\citenamefont {Zhu},
  \citenamefont {Gauthier}, \citenamefont {Morin}, \citenamefont {Wu},
  \citenamefont {Carmichael},\ and\ \citenamefont {Mossberg}}]{zhu1990vacuum}%
  \BibitemOpen
  \bibfield  {author} {\bibinfo {author} {\bibfnamefont {Y.}~\bibnamefont
  {Zhu}}, \bibinfo {author} {\bibfnamefont {D.~J.}\ \bibnamefont {Gauthier}},
  \bibinfo {author} {\bibfnamefont {S.~E.}\ \bibnamefont {Morin}}, \bibinfo
  {author} {\bibfnamefont {Q.}~\bibnamefont {Wu}}, \bibinfo {author}
  {\bibfnamefont {H.~J.}\ \bibnamefont {Carmichael}}, \ and\ \bibinfo {author}
  {\bibfnamefont {T.~W.}\ \bibnamefont {Mossberg}},\ }\href {\doibase
  10.1103/PhysRevLett.64.2499} {\bibfield  {journal} {\bibinfo  {journal}
  {Phys. Rev. Lett.}\ }\textbf {\bibinfo {volume} {64}},\ \bibinfo {pages}
  {2499} (\bibinfo {year} {1990})}\BibitemShut {NoStop}%
\bibitem [{\citenamefont {Thompson}\ \emph {et~al.}(1992)\citenamefont
  {Thompson}, \citenamefont {Rempe},\ and\ \citenamefont
  {Kimble}}]{thompson1992observation}%
  \BibitemOpen
  \bibfield  {author} {\bibinfo {author} {\bibfnamefont {R.~J.}\ \bibnamefont
  {Thompson}}, \bibinfo {author} {\bibfnamefont {G.}~\bibnamefont {Rempe}}, \
  and\ \bibinfo {author} {\bibfnamefont {H.~J.}\ \bibnamefont {Kimble}},\
  }\href {\doibase 10.1103/PhysRevLett.68.1132} {\bibfield  {journal} {\bibinfo
   {journal} {Phys. Rev. Lett.}\ }\textbf {\bibinfo {volume} {68}},\ \bibinfo
  {pages} {1132} (\bibinfo {year} {1992})}\BibitemShut {NoStop}%
\bibitem [{\citenamefont {Carmichael}\ \emph {et~al.}(1989)\citenamefont
  {Carmichael}, \citenamefont {Brecha}, \citenamefont {Raizen}, \citenamefont
  {Kimble},\ and\ \citenamefont {Rice}}]{carmichael1989subnatural}%
  \BibitemOpen
  \bibfield  {author} {\bibinfo {author} {\bibfnamefont {H.~J.}\ \bibnamefont
  {Carmichael}}, \bibinfo {author} {\bibfnamefont {R.~J.}\ \bibnamefont
  {Brecha}}, \bibinfo {author} {\bibfnamefont {M.~G.}\ \bibnamefont {Raizen}},
  \bibinfo {author} {\bibfnamefont {H.~J.}\ \bibnamefont {Kimble}}, \ and\
  \bibinfo {author} {\bibfnamefont {P.~R.}\ \bibnamefont {Rice}},\ }\href
  {\doibase 10.1103/PhysRevA.40.5516} {\bibfield  {journal} {\bibinfo
  {journal} {Phys. Rev. A}\ }\textbf {\bibinfo {volume} {40}},\ \bibinfo
  {pages} {5516} (\bibinfo {year} {1989})}\BibitemShut {NoStop}%
\bibitem [{\citenamefont {Carmichael}\ \emph {et~al.}(1991)\citenamefont
  {Carmichael}, \citenamefont {Brecha},\ and\ \citenamefont
  {Rice}}]{carmichael1991quantum}%
  \BibitemOpen
  \bibfield  {author} {\bibinfo {author} {\bibfnamefont {H.~J.}\ \bibnamefont
  {Carmichael}}, \bibinfo {author} {\bibfnamefont {R.~J.}\ \bibnamefont
  {Brecha}}, \ and\ \bibinfo {author} {\bibfnamefont {P.~R.}\ \bibnamefont
  {Rice}},\ }\href@noop {} {\bibfield  {journal} {\bibinfo  {journal} {Opt.
  Commun.}\ }\textbf {\bibinfo {volume} {82}},\ \bibinfo {pages} {73} (\bibinfo
  {year} {1991})}\BibitemShut {NoStop}%
\bibitem [{\citenamefont {Imamo\ifmmode~\breve{g}\else
  \u{g}\fi{}lu}(2009)}]{imamouglu2009cavity}%
  \BibitemOpen
  \bibfield  {author} {\bibinfo {author} {\bibfnamefont {A.}~\bibnamefont
  {Imamo\ifmmode~\breve{g}\else \u{g}\fi{}lu}},\ }\href {\doibase
  10.1103/PhysRevLett.102.083602} {\bibfield  {journal} {\bibinfo  {journal}
  {Phys. Rev. Lett.}\ }\textbf {\bibinfo {volume} {102}},\ \bibinfo {pages}
  {083602} (\bibinfo {year} {2009})}\BibitemShut {NoStop}%
\bibitem [{\citenamefont {Soykal}\ and\ \citenamefont
  {Flatt\'e}(2010)}]{soykal2010strong}%
  \BibitemOpen
  \bibfield  {author} {\bibinfo {author} {\bibfnamefont {O.~O.}\ \bibnamefont
  {Soykal}}\ and\ \bibinfo {author} {\bibfnamefont {M.~E.}\ \bibnamefont
  {Flatt\'e}},\ }\href {\doibase 10.1103/PhysRevLett.104.077202} {\bibfield
  {journal} {\bibinfo  {journal} {Phys. Rev. Lett.}\ }\textbf {\bibinfo
  {volume} {104}},\ \bibinfo {pages} {077202} (\bibinfo {year}
  {2010})}\BibitemShut {NoStop}%
\bibitem [{\citenamefont {Wesenberg}\ \emph {et~al.}(2009)\citenamefont
  {Wesenberg}, \citenamefont {Ardavan}, \citenamefont {Briggs}, \citenamefont
  {Morton}, \citenamefont {Schoelkopf}, \citenamefont {Schuster},\ and\
  \citenamefont {M\o{}lmer}}]{wesenberg2009quantum}%
  \BibitemOpen
  \bibfield  {author} {\bibinfo {author} {\bibfnamefont {J.~H.}\ \bibnamefont
  {Wesenberg}}, \bibinfo {author} {\bibfnamefont {A.}~\bibnamefont {Ardavan}},
  \bibinfo {author} {\bibfnamefont {G.~A.~D.}\ \bibnamefont {Briggs}}, \bibinfo
  {author} {\bibfnamefont {J.~J.~L.}\ \bibnamefont {Morton}}, \bibinfo {author}
  {\bibfnamefont {R.~J.}\ \bibnamefont {Schoelkopf}}, \bibinfo {author}
  {\bibfnamefont {D.~I.}\ \bibnamefont {Schuster}}, \ and\ \bibinfo {author}
  {\bibfnamefont {K.}~\bibnamefont {M\o{}lmer}},\ }\href {\doibase
  10.1103/PhysRevLett.103.070502} {\bibfield  {journal} {\bibinfo  {journal}
  {Phys. Rev. Lett.}\ }\textbf {\bibinfo {volume} {103}},\ \bibinfo {pages}
  {070502} (\bibinfo {year} {2009})}\BibitemShut {NoStop}%
\bibitem [{\citenamefont {Diniz}\ \emph {et~al.}(2011)\citenamefont {Diniz},
  \citenamefont {Portolan}, \citenamefont {Ferreira}, \citenamefont {G\'erard},
  \citenamefont {Bertet},\ and\ \citenamefont
  {Auff\`eves}}]{diniz2011strongly}%
  \BibitemOpen
  \bibfield  {author} {\bibinfo {author} {\bibfnamefont {I.}~\bibnamefont
  {Diniz}}, \bibinfo {author} {\bibfnamefont {S.}~\bibnamefont {Portolan}},
  \bibinfo {author} {\bibfnamefont {R.}~\bibnamefont {Ferreira}}, \bibinfo
  {author} {\bibfnamefont {J.~M.}\ \bibnamefont {G\'erard}}, \bibinfo {author}
  {\bibfnamefont {P.}~\bibnamefont {Bertet}}, \ and\ \bibinfo {author}
  {\bibfnamefont {A.}~\bibnamefont {Auff\`eves}},\ }\href {\doibase
  10.1103/PhysRevA.84.063810} {\bibfield  {journal} {\bibinfo  {journal} {Phys.
  Rev. A}\ }\textbf {\bibinfo {volume} {84}},\ \bibinfo {pages} {063810}
  (\bibinfo {year} {2011})}\BibitemShut {NoStop}%
\bibitem [{\citenamefont {Julsgaard}\ \emph {et~al.}(2013)\citenamefont
  {Julsgaard}, \citenamefont {Grezes}, \citenamefont {Bertet},\ and\
  \citenamefont {M\o{}lmer}}]{julsgaard2013quantum}%
  \BibitemOpen
  \bibfield  {author} {\bibinfo {author} {\bibfnamefont {B.}~\bibnamefont
  {Julsgaard}}, \bibinfo {author} {\bibfnamefont {C.}~\bibnamefont {Grezes}},
  \bibinfo {author} {\bibfnamefont {P.}~\bibnamefont {Bertet}}, \ and\ \bibinfo
  {author} {\bibfnamefont {K.}~\bibnamefont {M\o{}lmer}},\ }\href {\doibase
  10.1103/PhysRevLett.110.250503} {\bibfield  {journal} {\bibinfo  {journal}
  {Phys. Rev. Lett.}\ }\textbf {\bibinfo {volume} {110}},\ \bibinfo {pages}
  {250503} (\bibinfo {year} {2013})}\BibitemShut {NoStop}%
\bibitem [{\citenamefont {Schuster}\ \emph {et~al.}(2010)\citenamefont
  {Schuster}, \citenamefont {Sears}, \citenamefont {Ginossar}, \citenamefont
  {DiCarlo}, \citenamefont {Frunzio}, \citenamefont {Morton}, \citenamefont
  {Wu}, \citenamefont {Briggs}, \citenamefont {Buckley}, \citenamefont
  {Awschalom},\ and\ \citenamefont {Schoelkopf}}]{schuster2010high}%
  \BibitemOpen
  \bibfield  {author} {\bibinfo {author} {\bibfnamefont {D.~I.}\ \bibnamefont
  {Schuster}}, \bibinfo {author} {\bibfnamefont {A.~P.}\ \bibnamefont {Sears}},
  \bibinfo {author} {\bibfnamefont {E.}~\bibnamefont {Ginossar}}, \bibinfo
  {author} {\bibfnamefont {L.}~\bibnamefont {DiCarlo}}, \bibinfo {author}
  {\bibfnamefont {L.}~\bibnamefont {Frunzio}}, \bibinfo {author} {\bibfnamefont
  {J.~J.~L.}\ \bibnamefont {Morton}}, \bibinfo {author} {\bibfnamefont
  {H.}~\bibnamefont {Wu}}, \bibinfo {author} {\bibfnamefont {G.~A.~D.}\
  \bibnamefont {Briggs}}, \bibinfo {author} {\bibfnamefont {B.~B.}\
  \bibnamefont {Buckley}}, \bibinfo {author} {\bibfnamefont {D.~D.}\
  \bibnamefont {Awschalom}}, \ and\ \bibinfo {author} {\bibfnamefont {R.~J.}\
  \bibnamefont {Schoelkopf}},\ }\href {\doibase 10.1103/PhysRevLett.105.140501}
  {\bibfield  {journal} {\bibinfo  {journal} {Phys. Rev. Lett.}\ }\textbf
  {\bibinfo {volume} {105}},\ \bibinfo {pages} {140501} (\bibinfo {year}
  {2010})}\BibitemShut {NoStop}%
\bibitem [{\citenamefont {Kubo}\ \emph {et~al.}(2010)\citenamefont {Kubo},
  \citenamefont {Ong}, \citenamefont {Bertet}, \citenamefont {Vion},
  \citenamefont {Jacques}, \citenamefont {Zheng}, \citenamefont {Dr\'eau},
  \citenamefont {Roch}, \citenamefont {Auffeves}, \citenamefont {Jelezko},
  \citenamefont {Wrachtrup}, \citenamefont {Barthe}, \citenamefont {Bergonzo},\
  and\ \citenamefont {Esteve}}]{kubo2010strong}%
  \BibitemOpen
  \bibfield  {author} {\bibinfo {author} {\bibfnamefont {Y.}~\bibnamefont
  {Kubo}}, \bibinfo {author} {\bibfnamefont {F.~R.}\ \bibnamefont {Ong}},
  \bibinfo {author} {\bibfnamefont {P.}~\bibnamefont {Bertet}}, \bibinfo
  {author} {\bibfnamefont {D.}~\bibnamefont {Vion}}, \bibinfo {author}
  {\bibfnamefont {V.}~\bibnamefont {Jacques}}, \bibinfo {author} {\bibfnamefont
  {D.}~\bibnamefont {Zheng}}, \bibinfo {author} {\bibfnamefont
  {A.}~\bibnamefont {Dr\'eau}}, \bibinfo {author} {\bibfnamefont {J.-F.}\
  \bibnamefont {Roch}}, \bibinfo {author} {\bibfnamefont {A.}~\bibnamefont
  {Auffeves}}, \bibinfo {author} {\bibfnamefont {F.}~\bibnamefont {Jelezko}},
  \bibinfo {author} {\bibfnamefont {J.}~\bibnamefont {Wrachtrup}}, \bibinfo
  {author} {\bibfnamefont {M.~F.}\ \bibnamefont {Barthe}}, \bibinfo {author}
  {\bibfnamefont {P.}~\bibnamefont {Bergonzo}}, \ and\ \bibinfo {author}
  {\bibfnamefont {D.}~\bibnamefont {Esteve}},\ }\href {\doibase
  10.1103/PhysRevLett.105.140502} {\bibfield  {journal} {\bibinfo  {journal}
  {Phys. Rev. Lett.}\ }\textbf {\bibinfo {volume} {105}},\ \bibinfo {pages}
  {140502} (\bibinfo {year} {2010})}\BibitemShut {NoStop}%
\bibitem [{\citenamefont {Wu}\ \emph {et~al.}(2010)\citenamefont {Wu},
  \citenamefont {George}, \citenamefont {Wesenberg}, \citenamefont {M\o{}lmer},
  \citenamefont {Schuster}, \citenamefont {Schoelkopf}, \citenamefont {Itoh},
  \citenamefont {Ardavan}, \citenamefont {Morton},\ and\ \citenamefont
  {Briggs}}]{wu2010h}%
  \BibitemOpen
  \bibfield  {author} {\bibinfo {author} {\bibfnamefont {H.}~\bibnamefont
  {Wu}}, \bibinfo {author} {\bibfnamefont {R.~E.}\ \bibnamefont {George}},
  \bibinfo {author} {\bibfnamefont {J.~H.}\ \bibnamefont {Wesenberg}}, \bibinfo
  {author} {\bibfnamefont {K.}~\bibnamefont {M\o{}lmer}}, \bibinfo {author}
  {\bibfnamefont {D.~I.}\ \bibnamefont {Schuster}}, \bibinfo {author}
  {\bibfnamefont {R.~J.}\ \bibnamefont {Schoelkopf}}, \bibinfo {author}
  {\bibfnamefont {K.~M.}\ \bibnamefont {Itoh}}, \bibinfo {author}
  {\bibfnamefont {A.}~\bibnamefont {Ardavan}}, \bibinfo {author} {\bibfnamefont
  {J.~J.~L.}\ \bibnamefont {Morton}}, \ and\ \bibinfo {author} {\bibfnamefont
  {G.~A.~D.}\ \bibnamefont {Briggs}},\ }\href {\doibase
  10.1103/PhysRevLett.105.140503} {\bibfield  {journal} {\bibinfo  {journal}
  {Phys. Rev. Lett.}\ }\textbf {\bibinfo {volume} {105}},\ \bibinfo {pages}
  {140503} (\bibinfo {year} {2010})}\BibitemShut {NoStop}%
\bibitem [{\citenamefont {Ams\"uss}\ \emph {et~al.}(2011)\citenamefont
  {Ams\"uss}, \citenamefont {Koller}, \citenamefont {N\"obauer}, \citenamefont
  {Putz}, \citenamefont {Rotter}, \citenamefont {Sandner}, \citenamefont
  {Schneider}, \citenamefont {Schramb\"ock}, \citenamefont {Steinhauser},
  \citenamefont {Ritsch}, \citenamefont {Schmiedmayer},\ and\ \citenamefont
  {Majer}}]{amsuss2011cavity}%
  \BibitemOpen
  \bibfield  {author} {\bibinfo {author} {\bibfnamefont {R.}~\bibnamefont
  {Ams\"uss}}, \bibinfo {author} {\bibfnamefont {C.}~\bibnamefont {Koller}},
  \bibinfo {author} {\bibfnamefont {T.}~\bibnamefont {N\"obauer}}, \bibinfo
  {author} {\bibfnamefont {S.}~\bibnamefont {Putz}}, \bibinfo {author}
  {\bibfnamefont {S.}~\bibnamefont {Rotter}}, \bibinfo {author} {\bibfnamefont
  {K.}~\bibnamefont {Sandner}}, \bibinfo {author} {\bibfnamefont
  {S.}~\bibnamefont {Schneider}}, \bibinfo {author} {\bibfnamefont
  {M.}~\bibnamefont {Schramb\"ock}}, \bibinfo {author} {\bibfnamefont
  {G.}~\bibnamefont {Steinhauser}}, \bibinfo {author} {\bibfnamefont
  {H.}~\bibnamefont {Ritsch}}, \bibinfo {author} {\bibfnamefont
  {J.}~\bibnamefont {Schmiedmayer}}, \ and\ \bibinfo {author} {\bibfnamefont
  {J.}~\bibnamefont {Majer}},\ }\href {\doibase 10.1103/PhysRevLett.107.060502}
  {\bibfield  {journal} {\bibinfo  {journal} {Phys. Rev. Lett.}\ }\textbf
  {\bibinfo {volume} {107}},\ \bibinfo {pages} {060502} (\bibinfo {year}
  {2011})}\BibitemShut {NoStop}%
\bibitem [{\citenamefont {Abe}\ \emph {et~al.}(2011)\citenamefont {Abe},
  \citenamefont {Wu}, \citenamefont {Ardavan},\ and\ \citenamefont
  {Morton}}]{abe2011electron}%
  \BibitemOpen
  \bibfield  {author} {\bibinfo {author} {\bibfnamefont {E.}~\bibnamefont
  {Abe}}, \bibinfo {author} {\bibfnamefont {H.}~\bibnamefont {Wu}}, \bibinfo
  {author} {\bibfnamefont {A.}~\bibnamefont {Ardavan}}, \ and\ \bibinfo
  {author} {\bibfnamefont {J.~J.~L.}\ \bibnamefont {Morton}},\ }\href {\doibase
  10.1063/1.3601930} {\bibfield  {journal} {\bibinfo  {journal} {Appl. Phys.
  Lett.}\ }\textbf {\bibinfo {volume} {98}},\ \bibinfo {pages} {251108}
  (\bibinfo {year} {2011})}\BibitemShut {NoStop}%
\bibitem [{\citenamefont {Huebl}\ \emph {et~al.}(2013)\citenamefont {Huebl},
  \citenamefont {Zollitsch}, \citenamefont {Lotze}, \citenamefont {Hocke},
  \citenamefont {Greifenstein}, \citenamefont {Marx}, \citenamefont {Gross},\
  and\ \citenamefont {Goennenwein}}]{huebl2013high}%
  \BibitemOpen
  \bibfield  {author} {\bibinfo {author} {\bibfnamefont {H.}~\bibnamefont
  {Huebl}}, \bibinfo {author} {\bibfnamefont {C.~W.}\ \bibnamefont
  {Zollitsch}}, \bibinfo {author} {\bibfnamefont {J.}~\bibnamefont {Lotze}},
  \bibinfo {author} {\bibfnamefont {F.}~\bibnamefont {Hocke}}, \bibinfo
  {author} {\bibfnamefont {M.}~\bibnamefont {Greifenstein}}, \bibinfo {author}
  {\bibfnamefont {A.}~\bibnamefont {Marx}}, \bibinfo {author} {\bibfnamefont
  {R.}~\bibnamefont {Gross}}, \ and\ \bibinfo {author} {\bibfnamefont
  {S.~T.~B.}\ \bibnamefont {Goennenwein}},\ }\href {\doibase
  10.1103/PhysRevLett.111.127003} {\bibfield  {journal} {\bibinfo  {journal}
  {Phys. Rev. Lett.}\ }\textbf {\bibinfo {volume} {111}},\ \bibinfo {pages}
  {127003} (\bibinfo {year} {2013})}\BibitemShut {NoStop}%
\bibitem [{\citenamefont {Goryachev}\ \emph {et~al.}(2014)\citenamefont
  {Goryachev}, \citenamefont {Farr}, \citenamefont {Creedon}, \citenamefont
  {Fan}, \citenamefont {Kostylev},\ and\ \citenamefont
  {Tobar}}]{goryachev2014high}%
  \BibitemOpen
  \bibfield  {author} {\bibinfo {author} {\bibfnamefont {M.}~\bibnamefont
  {Goryachev}}, \bibinfo {author} {\bibfnamefont {W.~G.}\ \bibnamefont {Farr}},
  \bibinfo {author} {\bibfnamefont {D.~L.}\ \bibnamefont {Creedon}}, \bibinfo
  {author} {\bibfnamefont {Y.}~\bibnamefont {Fan}}, \bibinfo {author}
  {\bibfnamefont {M.}~\bibnamefont {Kostylev}}, \ and\ \bibinfo {author}
  {\bibfnamefont {M.~E.}\ \bibnamefont {Tobar}},\ }\href {\doibase
  10.1103/PhysRevApplied.2.054002} {\bibfield  {journal} {\bibinfo  {journal}
  {Phys. Rev. Appl.}\ }\textbf {\bibinfo {volume} {2}},\ \bibinfo {pages}
  {054002} (\bibinfo {year} {2014})}\BibitemShut {NoStop}%
\bibitem [{\citenamefont {Tabuchi}\ \emph {et~al.}(2014)\citenamefont
  {Tabuchi}, \citenamefont {Ishino}, \citenamefont {Ishikawa}, \citenamefont
  {Yamazaki}, \citenamefont {Usami},\ and\ \citenamefont
  {Nakamura}}]{tabuchi2014tabuchi}%
  \BibitemOpen
  \bibfield  {author} {\bibinfo {author} {\bibfnamefont {Y.}~\bibnamefont
  {Tabuchi}}, \bibinfo {author} {\bibfnamefont {S.}~\bibnamefont {Ishino}},
  \bibinfo {author} {\bibfnamefont {T.}~\bibnamefont {Ishikawa}}, \bibinfo
  {author} {\bibfnamefont {R.}~\bibnamefont {Yamazaki}}, \bibinfo {author}
  {\bibfnamefont {K.}~\bibnamefont {Usami}}, \ and\ \bibinfo {author}
  {\bibfnamefont {Y.}~\bibnamefont {Nakamura}},\ }\href {\doibase
  10.1103/PhysRevLett.113.083603} {\bibfield  {journal} {\bibinfo  {journal}
  {Phys. Rev. Lett.}\ }\textbf {\bibinfo {volume} {113}},\ \bibinfo {pages}
  {083603} (\bibinfo {year} {2014})}\BibitemShut {NoStop}%
\bibitem [{\citenamefont {Zhang}\ \emph {et~al.}(2014)\citenamefont {Zhang},
  \citenamefont {Zou}, \citenamefont {Jiang},\ and\ \citenamefont
  {Tang}}]{zhang2014strongly}%
  \BibitemOpen
  \bibfield  {author} {\bibinfo {author} {\bibfnamefont {X.}~\bibnamefont
  {Zhang}}, \bibinfo {author} {\bibfnamefont {C.-L.}\ \bibnamefont {Zou}},
  \bibinfo {author} {\bibfnamefont {L.}~\bibnamefont {Jiang}}, \ and\ \bibinfo
  {author} {\bibfnamefont {H.~X.}\ \bibnamefont {Tang}},\ }\href {\doibase
  10.1103/PhysRevLett.113.156401} {\bibfield  {journal} {\bibinfo  {journal}
  {Phys. Rev. Lett.}\ }\textbf {\bibinfo {volume} {113}},\ \bibinfo {pages}
  {156401} (\bibinfo {year} {2014})}\BibitemShut {NoStop}%
\bibitem [{\citenamefont {Abdurakhimov}\ \emph {et~al.}(2015)\citenamefont
  {Abdurakhimov}, \citenamefont {Bunkov},\ and\ \citenamefont
  {Konstantinov}}]{abdurakhimov2015normal}%
  \BibitemOpen
  \bibfield  {author} {\bibinfo {author} {\bibfnamefont {L.~V.}\ \bibnamefont
  {Abdurakhimov}}, \bibinfo {author} {\bibfnamefont {Y.~M.}\ \bibnamefont
  {Bunkov}}, \ and\ \bibinfo {author} {\bibfnamefont {D.}~\bibnamefont
  {Konstantinov}},\ }\href {\doibase 10.1103/PhysRevLett.114.226402} {\bibfield
   {journal} {\bibinfo  {journal} {Phys. Rev. Lett.}\ }\textbf {\bibinfo
  {volume} {114}},\ \bibinfo {pages} {226402} (\bibinfo {year}
  {2015})}\BibitemShut {NoStop}%
\bibitem [{\citenamefont {Dini}\ \emph {et~al.}(2003)\citenamefont {Dini},
  \citenamefont {Kohler}, \citenamefont {Tredicucci}, \citenamefont {Biasiol},\
  and\ \citenamefont {Sorba}}]{dini2003prl}%
  \BibitemOpen
  \bibfield  {author} {\bibinfo {author} {\bibfnamefont {D.}~\bibnamefont
  {Dini}}, \bibinfo {author} {\bibfnamefont {R.}~\bibnamefont {Kohler}},
  \bibinfo {author} {\bibfnamefont {A.}~\bibnamefont {Tredicucci}}, \bibinfo
  {author} {\bibfnamefont {G.}~\bibnamefont {Biasiol}}, \ and\ \bibinfo
  {author} {\bibfnamefont {L.}~\bibnamefont {Sorba}},\ }\href@noop {}
  {\bibfield  {journal} {\bibinfo  {journal} {Phys. Rev. Lett.}\ }\textbf
  {\bibinfo {volume} {90}},\ \bibinfo {pages} {116401} (\bibinfo {year}
  {2003})}\BibitemShut {NoStop}%
\bibitem [{\citenamefont {Sapienza}\ \emph {et~al.}(2008)\citenamefont
  {Sapienza}, \citenamefont {Vasanelli}, \citenamefont {Colombelli},
  \citenamefont {Ciuti}, \citenamefont {Chassagneux}, \citenamefont {Manquest},
  \citenamefont {Gennser},\ and\ \citenamefont {Sistori}}]{sapienza2008prl}%
  \BibitemOpen
  \bibfield  {author} {\bibinfo {author} {\bibfnamefont {L.}~\bibnamefont
  {Sapienza}}, \bibinfo {author} {\bibfnamefont {A.}~\bibnamefont {Vasanelli}},
  \bibinfo {author} {\bibfnamefont {R.}~\bibnamefont {Colombelli}}, \bibinfo
  {author} {\bibfnamefont {C.}~\bibnamefont {Ciuti}}, \bibinfo {author}
  {\bibfnamefont {Y.}~\bibnamefont {Chassagneux}}, \bibinfo {author}
  {\bibfnamefont {C.}~\bibnamefont {Manquest}}, \bibinfo {author}
  {\bibfnamefont {U.}~\bibnamefont {Gennser}}, \ and\ \bibinfo {author}
  {\bibfnamefont {C.}~\bibnamefont {Sistori}},\ }\href@noop {} {\bibfield
  {journal} {\bibinfo  {journal} {Phys. Rev. Lett.}\ }\textbf {\bibinfo
  {volume} {100}},\ \bibinfo {pages} {136806} (\bibinfo {year}
  {2008})}\BibitemShut {NoStop}%
\bibitem [{\citenamefont {Todorov}\ \emph {et~al.}(2010)\citenamefont
  {Todorov}, \citenamefont {Anrews}, \citenamefont {Colombelli}, \citenamefont
  {De~Liberato}, \citenamefont {Ciuti}, \citenamefont {Lkang}, \citenamefont
  {Strasser},\ and\ \citenamefont {C.}}]{todorov2010prl}%
  \BibitemOpen
  \bibfield  {author} {\bibinfo {author} {\bibfnamefont {Y.}~\bibnamefont
  {Todorov}}, \bibinfo {author} {\bibfnamefont {A.~M.}\ \bibnamefont {Anrews}},
  \bibinfo {author} {\bibfnamefont {R.}~\bibnamefont {Colombelli}}, \bibinfo
  {author} {\bibfnamefont {S.}~\bibnamefont {De~Liberato}}, \bibinfo {author}
  {\bibfnamefont {C.}~\bibnamefont {Ciuti}}, \bibinfo {author} {\bibfnamefont
  {P.}~\bibnamefont {Lkang}}, \bibinfo {author} {\bibfnamefont
  {G.}~\bibnamefont {Strasser}}, \ and\ \bibinfo {author} {\bibfnamefont
  {S.}~\bibnamefont {C.}},\ }\href@noop {} {\bibfield  {journal} {\bibinfo
  {journal} {Phys. Rev. Lett.}\ }\textbf {\bibinfo {volume} {105}},\ \bibinfo
  {pages} {196402} (\bibinfo {year} {2010})}\BibitemShut {NoStop}%
\bibitem [{\citenamefont {Todorov}(2014)}]{todorov2014prb}%
  \BibitemOpen
  \bibfield  {author} {\bibinfo {author} {\bibfnamefont {Y.}~\bibnamefont
  {Todorov}},\ }\href@noop {} {\bibfield  {journal} {\bibinfo  {journal} {Phys.
  Rev. B}\ }\textbf {\bibinfo {volume} {89}},\ \bibinfo {pages} {075115}
  (\bibinfo {year} {2014})}\BibitemShut {NoStop}%
\bibitem [{\citenamefont {Todorov}(2015)}]{todorov2015prb}%
  \BibitemOpen
  \bibfield  {author} {\bibinfo {author} {\bibfnamefont {Y.}~\bibnamefont
  {Todorov}},\ }\href@noop {} {\bibfield  {journal} {\bibinfo  {journal} {Phys.
  Rev. B}\ }\textbf {\bibinfo {volume} {91}},\ \bibinfo {pages} {125409}
  (\bibinfo {year} {2015})}\BibitemShut {NoStop}%
\bibitem [{\citenamefont {Muravev}\ \emph {et~al.}(2011)\citenamefont
  {Muravev}, \citenamefont {Andreev}, \citenamefont {Kukushkin}, \citenamefont
  {Schmult},\ and\ \citenamefont {Dietsche}}]{muravev2011prb}%
  \BibitemOpen
  \bibfield  {author} {\bibinfo {author} {\bibfnamefont {V.~M.}\ \bibnamefont
  {Muravev}}, \bibinfo {author} {\bibfnamefont {I.~V.}\ \bibnamefont
  {Andreev}}, \bibinfo {author} {\bibfnamefont {I.~V.}\ \bibnamefont
  {Kukushkin}}, \bibinfo {author} {\bibfnamefont {S.}~\bibnamefont {Schmult}},
  \ and\ \bibinfo {author} {\bibfnamefont {W.}~\bibnamefont {Dietsche}},\
  }\href {\doibase 10.1103/PhysRevB.83.075309} {\bibfield  {journal} {\bibinfo
  {journal} {Phys. Rev. B}\ }\textbf {\bibinfo {volume} {83}},\ \bibinfo
  {pages} {075309} (\bibinfo {year} {2011})}\BibitemShut {NoStop}%
\bibitem [{\citenamefont {Muravev}\ \emph {et~al.}(2013)\citenamefont
  {Muravev}, \citenamefont {Gusikhin}, \citenamefont {Andreev},\ and\
  \citenamefont {Kukushkin}}]{muravev2013prb}%
  \BibitemOpen
  \bibfield  {author} {\bibinfo {author} {\bibfnamefont {V.~M.}\ \bibnamefont
  {Muravev}}, \bibinfo {author} {\bibfnamefont {P.~A.}\ \bibnamefont
  {Gusikhin}}, \bibinfo {author} {\bibfnamefont {I.~V.}\ \bibnamefont
  {Andreev}}, \ and\ \bibinfo {author} {\bibfnamefont {I.~V.}\ \bibnamefont
  {Kukushkin}},\ }\href {\doibase 10.1103/PhysRevB.87.045307} {\bibfield
  {journal} {\bibinfo  {journal} {Phys. Rev. B}\ }\textbf {\bibinfo {volume}
  {87}},\ \bibinfo {pages} {045307} (\bibinfo {year} {2013})}\BibitemShut
  {NoStop}%
\bibitem [{\citenamefont {Shikin}(2002)}]{shikin2002}%
  \BibitemOpen
  \bibfield  {author} {\bibinfo {author} {\bibfnamefont {V.~B.}\ \bibnamefont
  {Shikin}},\ }\href@noop {} {\bibfield  {journal} {\bibinfo  {journal} {JETP
  Letters}\ }\textbf {\bibinfo {volume} {75}},\ \bibinfo {pages} {29} (\bibinfo
  {year} {2002})}\BibitemShut {NoStop}%
\bibitem [{\citenamefont {Ciuti}\ \emph {et~al.}(2005)\citenamefont {Ciuti},
  \citenamefont {Bastard},\ and\ \citenamefont {Carusotto}}]{ciuti2005prb}%
  \BibitemOpen
  \bibfield  {author} {\bibinfo {author} {\bibfnamefont {C.}~\bibnamefont
  {Ciuti}}, \bibinfo {author} {\bibfnamefont {G.}~\bibnamefont {Bastard}}, \
  and\ \bibinfo {author} {\bibfnamefont {I.}~\bibnamefont {Carusotto}},\
  }\href@noop {} {\bibfield  {journal} {\bibinfo  {journal} {Phys. Rev. B}\
  }\textbf {\bibinfo {volume} {72}},\ \bibinfo {pages} {115303} (\bibinfo
  {year} {2005})}\BibitemShut {NoStop}%
\bibitem [{\citenamefont {Hagenm\"uller}\ \emph {et~al.}(2010)\citenamefont
  {Hagenm\"uller}, \citenamefont {De~Liberato},\ and\ \citenamefont
  {Ciuti}}]{hagenmuller2010ultrastrong}%
  \BibitemOpen
  \bibfield  {author} {\bibinfo {author} {\bibfnamefont {D.}~\bibnamefont
  {Hagenm\"uller}}, \bibinfo {author} {\bibfnamefont {S.}~\bibnamefont
  {De~Liberato}}, \ and\ \bibinfo {author} {\bibfnamefont {C.}~\bibnamefont
  {Ciuti}},\ }\href {\doibase 10.1103/PhysRevB.81.235303} {\bibfield  {journal}
  {\bibinfo  {journal} {Phys. Rev. B}\ }\textbf {\bibinfo {volume} {81}},\
  \bibinfo {pages} {235303} (\bibinfo {year} {2010})}\BibitemShut {NoStop}%
\bibitem [{\citenamefont {Hagenmuller}\ and\ \citenamefont
  {Ciuti}(2012)}]{hagenmuller2012prl}%
  \BibitemOpen
  \bibfield  {author} {\bibinfo {author} {\bibfnamefont {D.}~\bibnamefont
  {Hagenmuller}}\ and\ \bibinfo {author} {\bibfnamefont {C.}~\bibnamefont
  {Ciuti}},\ }\href {\doibase 10.1103/PhysRevLett.109.267403} {\bibfield
  {journal} {\bibinfo  {journal} {Phys. Rev. Lett.}\ }\textbf {\bibinfo
  {volume} {109}},\ \bibinfo {pages} {267403} (\bibinfo {year}
  {2012})}\BibitemShut {NoStop}%
\bibitem [{\citenamefont {Chirolli}\ \emph {et~al.}(2012)\citenamefont
  {Chirolli}, \citenamefont {Polini}, \citenamefont {Giovannetti},\ and\
  \citenamefont {MacDonald}}]{chirolli2012prl}%
  \BibitemOpen
  \bibfield  {author} {\bibinfo {author} {\bibfnamefont {L.}~\bibnamefont
  {Chirolli}}, \bibinfo {author} {\bibfnamefont {M.}~\bibnamefont {Polini}},
  \bibinfo {author} {\bibfnamefont {V.}~\bibnamefont {Giovannetti}}, \ and\
  \bibinfo {author} {\bibfnamefont {A.~H.}\ \bibnamefont {MacDonald}},\ }\href
  {\doibase 10.1103/PhysRevLett.109.267404} {\bibfield  {journal} {\bibinfo
  {journal} {Phys. Rev. Lett.}\ }\textbf {\bibinfo {volume} {109}},\ \bibinfo
  {pages} {267404} (\bibinfo {year} {2012})}\BibitemShut {NoStop}%
\bibitem [{\citenamefont {Scalari}\ \emph {et~al.}(2012)\citenamefont
  {Scalari}, \citenamefont {Maissen}, \citenamefont {Tur{\v c}inkov{\'a}},
  \citenamefont {Hagenm{\"u}ller}, \citenamefont {De~Liberato}, \citenamefont
  {Ciuti}, \citenamefont {Reichl}, \citenamefont {Schuh}, \citenamefont
  {Wegscheider}, \citenamefont {Beck},\ and\ \citenamefont
  {Faist}}]{scalari2012ultrastrong}%
  \BibitemOpen
  \bibfield  {author} {\bibinfo {author} {\bibfnamefont {G.}~\bibnamefont
  {Scalari}}, \bibinfo {author} {\bibfnamefont {C.}~\bibnamefont {Maissen}},
  \bibinfo {author} {\bibfnamefont {D.}~\bibnamefont {Tur{\v c}inkov{\'a}}},
  \bibinfo {author} {\bibfnamefont {D.}~\bibnamefont {Hagenm{\"u}ller}},
  \bibinfo {author} {\bibfnamefont {S.}~\bibnamefont {De~Liberato}}, \bibinfo
  {author} {\bibfnamefont {C.}~\bibnamefont {Ciuti}}, \bibinfo {author}
  {\bibfnamefont {C.}~\bibnamefont {Reichl}}, \bibinfo {author} {\bibfnamefont
  {D.}~\bibnamefont {Schuh}}, \bibinfo {author} {\bibfnamefont
  {W.}~\bibnamefont {Wegscheider}}, \bibinfo {author} {\bibfnamefont
  {M.}~\bibnamefont {Beck}}, \ and\ \bibinfo {author} {\bibfnamefont
  {J.}~\bibnamefont {Faist}},\ }\href {\doibase 10.1126/science.1216022}
  {\bibfield  {journal} {\bibinfo  {journal} {Science}\ }\textbf {\bibinfo
  {volume} {335}},\ \bibinfo {pages} {1323} (\bibinfo {year}
  {2012})}\BibitemShut {NoStop}%
\bibitem [{\citenamefont {Zhang}\ \emph {et~al.}(2016)\citenamefont {Zhang},
  \citenamefont {Lou}, \citenamefont {Li}, \citenamefont {Reno}, \citenamefont
  {Pan}, \citenamefont {Watson}, \citenamefont {Manfra},\ and\ \citenamefont
  {Kono}}]{zhang2016collective}%
  \BibitemOpen
  \bibfield  {author} {\bibinfo {author} {\bibfnamefont {Q.}~\bibnamefont
  {Zhang}}, \bibinfo {author} {\bibfnamefont {M.}~\bibnamefont {Lou}}, \bibinfo
  {author} {\bibfnamefont {X.}~\bibnamefont {Li}}, \bibinfo {author}
  {\bibfnamefont {J.~L.}\ \bibnamefont {Reno}}, \bibinfo {author}
  {\bibfnamefont {W.}~\bibnamefont {Pan}}, \bibinfo {author} {\bibfnamefont
  {J.~D.}\ \bibnamefont {Watson}}, \bibinfo {author} {\bibfnamefont {M.~J.}\
  \bibnamefont {Manfra}}, \ and\ \bibinfo {author} {\bibfnamefont
  {J.}~\bibnamefont {Kono}},\ }\href@noop {} {\bibfield  {journal} {\bibinfo
  {journal} {Nat. Phys.}\ }\textbf {\bibinfo {volume} {12}},\ \bibinfo {pages}
  {1005} (\bibinfo {year} {2016})}\BibitemShut {NoStop}%
\bibitem [{\citenamefont {Li}\ \emph {et~al.}(2018)\citenamefont {Li},
  \citenamefont {Bamba}, \citenamefont {Zhang}, \citenamefont {Fallahi},
  \citenamefont {Garden}, \citenamefont {Gao}, \citenamefont {Lou},
  \citenamefont {Manfra},\ and\ \citenamefont {Kono}}]{li2018bloch}%
  \BibitemOpen
  \bibfield  {author} {\bibinfo {author} {\bibfnamefont {X.}~\bibnamefont
  {Li}}, \bibinfo {author} {\bibfnamefont {M.}~\bibnamefont {Bamba}}, \bibinfo
  {author} {\bibfnamefont {Q.}~\bibnamefont {Zhang}}, \bibinfo {author}
  {\bibfnamefont {S.}~\bibnamefont {Fallahi}}, \bibinfo {author} {\bibfnamefont
  {G.~C.}\ \bibnamefont {Garden}}, \bibinfo {author} {\bibfnamefont
  {W.}~\bibnamefont {Gao}}, \bibinfo {author} {\bibfnamefont {K.}~\bibnamefont
  {Lou}, \bibfnamefont {M.~Yoshioka}}, \bibinfo {author} {\bibfnamefont
  {M.~J.}\ \bibnamefont {Manfra}}, \ and\ \bibinfo {author} {\bibfnamefont
  {J.}~\bibnamefont {Kono}},\ }\href@noop {} {\bibfield  {journal} {\bibinfo
  {journal} {Nat. Photonics}\ }\textbf {\bibinfo {volume} {12}},\ \bibinfo
  {pages} {324} (\bibinfo {year} {2018})}\BibitemShut {NoStop}%
\bibitem [{\citenamefont {Bartolo}\ and\ \citenamefont
  {Ciuti}(2018)}]{bartolo2018prb}%
  \BibitemOpen
  \bibfield  {author} {\bibinfo {author} {\bibfnamefont {N.}~\bibnamefont
  {Bartolo}}\ and\ \bibinfo {author} {\bibfnamefont {C.}~\bibnamefont
  {Ciuti}},\ }\href@noop {} {\bibfield  {journal} {\bibinfo  {journal} {Phys.
  Rev. B}\ }\textbf {\bibinfo {volume} {98}},\ \bibinfo {pages} {205301}
  (\bibinfo {year} {2018})}\BibitemShut {NoStop}%
\bibitem [{\citenamefont {Paravicini-Bagliani}\ \emph
  {et~al.}(2018)\citenamefont {Paravicini-Bagliani}, \citenamefont
  {Appugliese}, \citenamefont {Richter}, \citenamefont {Valmorra},
  \citenamefont {Keller}, \citenamefont {Beck}, \citenamefont {Bartolo},
  \citenamefont {Rossler}, \citenamefont {Ihn}, \citenamefont {Ensslin},
  \citenamefont {Ciuti}, \citenamefont {Scalari},\ and\ \citenamefont
  {Faist}}]{paravicini2018arXiv}%
  \BibitemOpen
  \bibfield  {author} {\bibinfo {author} {\bibfnamefont {G.~L.}\ \bibnamefont
  {Paravicini-Bagliani}}, \bibinfo {author} {\bibfnamefont {F.}~\bibnamefont
  {Appugliese}}, \bibinfo {author} {\bibfnamefont {E.}~\bibnamefont {Richter}},
  \bibinfo {author} {\bibfnamefont {F.}~\bibnamefont {Valmorra}}, \bibinfo
  {author} {\bibfnamefont {J.}~\bibnamefont {Keller}}, \bibinfo {author}
  {\bibfnamefont {M.}~\bibnamefont {Beck}}, \bibinfo {author} {\bibfnamefont
  {N.}~\bibnamefont {Bartolo}}, \bibinfo {author} {\bibfnamefont
  {C.}~\bibnamefont {Rossler}}, \bibinfo {author} {\bibfnamefont
  {T.}~\bibnamefont {Ihn}}, \bibinfo {author} {\bibfnamefont {K.}~\bibnamefont
  {Ensslin}}, \bibinfo {author} {\bibfnamefont {C.}~\bibnamefont {Ciuti}},
  \bibinfo {author} {\bibfnamefont {G.}~\bibnamefont {Scalari}}, \ and\
  \bibinfo {author} {\bibfnamefont {J.}~\bibnamefont {Faist}},\ }\href@noop {}
  {\bibfield  {journal} {\bibinfo  {journal} {arXiv preprint arXiv:1805.00846}\
  } (\bibinfo {year} {2018})}\BibitemShut {NoStop}%
\bibitem [{\citenamefont {Keller}\ \emph {et~al.}(2018)\citenamefont {Keller},
  \citenamefont {Scalari}, \citenamefont {Appugliese}, \citenamefont
  {Rajabali}, \citenamefont {Maissen}, \citenamefont {Beck}, \citenamefont
  {Haase}, \citenamefont {Lehner}, \citenamefont {Wegscheider}, \citenamefont
  {Failla}, \citenamefont {Myronov}, \citenamefont {Leadley}, \citenamefont
  {Natal},\ and\ \citenamefont {Faist}}]{keller2018arXiv}%
  \BibitemOpen
  \bibfield  {author} {\bibinfo {author} {\bibfnamefont {J.}~\bibnamefont
  {Keller}}, \bibinfo {author} {\bibfnamefont {G.}~\bibnamefont {Scalari}},
  \bibinfo {author} {\bibfnamefont {F.}~\bibnamefont {Appugliese}}, \bibinfo
  {author} {\bibfnamefont {S.}~\bibnamefont {Rajabali}}, \bibinfo {author}
  {\bibfnamefont {C.}~\bibnamefont {Maissen}}, \bibinfo {author} {\bibfnamefont
  {M.}~\bibnamefont {Beck}}, \bibinfo {author} {\bibfnamefont {J.}~\bibnamefont
  {Haase}}, \bibinfo {author} {\bibfnamefont {C.~A.}\ \bibnamefont {Lehner}},
  \bibinfo {author} {\bibfnamefont {W.}~\bibnamefont {Wegscheider}}, \bibinfo
  {author} {\bibfnamefont {M.}~\bibnamefont {Failla}}, \bibinfo {author}
  {\bibfnamefont {M.}~\bibnamefont {Myronov}}, \bibinfo {author} {\bibfnamefont
  {J.}~\bibnamefont {Leadley}, \bibfnamefont {David R. Lloyd-Hughes}}, \bibinfo
  {author} {\bibfnamefont {P.}~\bibnamefont {Natal}}, \ and\ \bibinfo {author}
  {\bibfnamefont {J.}~\bibnamefont {Faist}},\ }\href@noop {} {\bibfield
  {journal} {\bibinfo  {journal} {arXiv preprint arXiv:1708.07773}\ } (\bibinfo
  {year} {2018})}\BibitemShut {NoStop}%
\bibitem [{\citenamefont {Abdurakhimov}\ \emph {et~al.}(2016)\citenamefont
  {Abdurakhimov}, \citenamefont {Yamashiro}, \citenamefont {Badrutdinov},\ and\
  \citenamefont {Konstantinov}}]{abdurakhimov2016strong}%
  \BibitemOpen
  \bibfield  {author} {\bibinfo {author} {\bibfnamefont {L.~V.}\ \bibnamefont
  {Abdurakhimov}}, \bibinfo {author} {\bibfnamefont {R.}~\bibnamefont
  {Yamashiro}}, \bibinfo {author} {\bibfnamefont {A.~O.}\ \bibnamefont
  {Badrutdinov}}, \ and\ \bibinfo {author} {\bibfnamefont {D.}~\bibnamefont
  {Konstantinov}},\ }\href {\doibase 10.1103/PhysRevLett.117.056803} {\bibfield
   {journal} {\bibinfo  {journal} {Phys. Rev. Lett.}\ }\textbf {\bibinfo
  {volume} {117}},\ \bibinfo {pages} {056803} (\bibinfo {year}
  {2016})}\BibitemShut {NoStop}%
\bibitem [{\citenamefont {Kogelnik}\ and\ \citenamefont
  {Li}(1966)}]{kogelnik1966laser}%
  \BibitemOpen
  \bibfield  {author} {\bibinfo {author} {\bibfnamefont {H.}~\bibnamefont
  {Kogelnik}}\ and\ \bibinfo {author} {\bibfnamefont {T.}~\bibnamefont {Li}},\
  }\href@noop {} {\bibfield  {journal} {\bibinfo  {journal} {Appl. Opt.}\
  }\textbf {\bibinfo {volume} {5}},\ \bibinfo {pages} {1550} (\bibinfo {year}
  {1966})}\BibitemShut {NoStop}%
\bibitem [{\citenamefont {Monarkha}\ and\ \citenamefont
  {Kono}(2013)}]{monarkha2013two}%
  \BibitemOpen
  \bibfield  {author} {\bibinfo {author} {\bibfnamefont {Y.}~\bibnamefont
  {Monarkha}}\ and\ \bibinfo {author} {\bibfnamefont {K.}~\bibnamefont
  {Kono}},\ }\href@noop {} {\emph {\bibinfo {title} {Two-dimensional Coulomb
  liquids and solids}}}\ (\bibinfo  {publisher} {Springer-Verlag, Berlin},\
  \bibinfo {year} {2013})\BibitemShut {NoStop}%
\bibitem [{\citenamefont {Julsgaard}\ and\ \citenamefont
  {M{\o}lmer}(2012)}]{julsgaard2012dynamical}%
  \BibitemOpen
  \bibfield  {author} {\bibinfo {author} {\bibfnamefont {B.}~\bibnamefont
  {Julsgaard}}\ and\ \bibinfo {author} {\bibfnamefont {K.}~\bibnamefont
  {M{\o}lmer}},\ }\href@noop {} {\bibfield  {journal} {\bibinfo  {journal}
  {Phys. Rev. A}\ }\textbf {\bibinfo {volume} {86}},\ \bibinfo {pages} {063810}
  (\bibinfo {year} {2012})}\BibitemShut {NoStop}%
\bibitem [{\citenamefont {Meystre}\ and\ \citenamefont
  {Murray}(2007)}]{meystre2007quantum}%
  \BibitemOpen
  \bibfield  {author} {\bibinfo {author} {\bibfnamefont {P.}~\bibnamefont
  {Meystre}}\ and\ \bibinfo {author} {\bibfnamefont {S.~I.}\ \bibnamefont
  {Murray}},\ }\href@noop {} {\emph {\bibinfo {title} {Elements of quantum
  optics}}}\ (\bibinfo  {publisher} {Springer-Verlag, Berlin},\ \bibinfo {year}
  {2007})\BibitemShut {NoStop}%
\bibitem [{\citenamefont {Walls}\ and\ \citenamefont
  {Milburn}(2008)}]{walls2008quantum}%
  \BibitemOpen
  \bibfield  {author} {\bibinfo {author} {\bibfnamefont {D.~F.}\ \bibnamefont
  {Walls}}\ and\ \bibinfo {author} {\bibfnamefont {G.~J.}\ \bibnamefont
  {Milburn}},\ }\href@noop {} {\emph {\bibinfo {title} {Quantum optics}}}\
  (\bibinfo  {publisher} {Springer-Verlag, Berlin},\ \bibinfo {year}
  {2008})\BibitemShut {NoStop}%
\bibitem [{\citenamefont {Collett}\ and\ \citenamefont
  {Gardiner}(1984)}]{collett1984squeezing}%
  \BibitemOpen
  \bibfield  {author} {\bibinfo {author} {\bibfnamefont {M.~J.}\ \bibnamefont
  {Collett}}\ and\ \bibinfo {author} {\bibfnamefont {C.~W.}\ \bibnamefont
  {Gardiner}},\ }\href {\doibase 10.1103/PhysRevA.30.1386} {\bibfield
  {journal} {\bibinfo  {journal} {Phys. Rev. A}\ }\textbf {\bibinfo {volume}
  {30}},\ \bibinfo {pages} {1386} (\bibinfo {year} {1984})}\BibitemShut
  {NoStop}%
\bibitem [{\citenamefont {Zadorozhko}\ \emph {et~al.}(2018)\citenamefont
  {Zadorozhko}, \citenamefont {Monarkha},\ and\ \citenamefont
  {Konstantinov}}]{zador2018polar}%
  \BibitemOpen
  \bibfield  {author} {\bibinfo {author} {\bibfnamefont {A.~A.}\ \bibnamefont
  {Zadorozhko}}, \bibinfo {author} {\bibfnamefont {Y.~P.}\ \bibnamefont
  {Monarkha}}, \ and\ \bibinfo {author} {\bibfnamefont {D.}~\bibnamefont
  {Konstantinov}},\ }\href {\doibase 10.1103/PhysRevLett.120.046802} {\bibfield
   {journal} {\bibinfo  {journal} {Phys. Rev. Lett.}\ }\textbf {\bibinfo
  {volume} {120}},\ \bibinfo {pages} {046802} (\bibinfo {year}
  {2018})}\BibitemShut {NoStop}%
\bibitem [{\citenamefont {Bai}\ \emph {et~al.}(2015)\citenamefont {Bai},
  \citenamefont {Harder}, \citenamefont {Chen}, \citenamefont {Fan},
  \citenamefont {Xiao},\ and\ \citenamefont {Hu}}]{bai2015spin}%
  \BibitemOpen
  \bibfield  {author} {\bibinfo {author} {\bibfnamefont {L.}~\bibnamefont
  {Bai}}, \bibinfo {author} {\bibfnamefont {M.}~\bibnamefont {Harder}},
  \bibinfo {author} {\bibfnamefont {Y.~P.}\ \bibnamefont {Chen}}, \bibinfo
  {author} {\bibfnamefont {X.}~\bibnamefont {Fan}}, \bibinfo {author}
  {\bibfnamefont {J.~Q.}\ \bibnamefont {Xiao}}, \ and\ \bibinfo {author}
  {\bibfnamefont {C.-M.}\ \bibnamefont {Hu}},\ }\href {\doibase
  10.1103/PhysRevLett.114.227201} {\bibfield  {journal} {\bibinfo  {journal}
  {Phys. Rev. Lett.}\ }\textbf {\bibinfo {volume} {114}},\ \bibinfo {pages}
  {227201} (\bibinfo {year} {2015})}\BibitemShut {NoStop}%
\bibitem [{\citenamefont {Todorov}\ and\ \citenamefont
  {Sirtoti}(2012)}]{todorov2012prb}%
  \BibitemOpen
  \bibfield  {author} {\bibinfo {author} {\bibfnamefont {Y.}~\bibnamefont
  {Todorov}}\ and\ \bibinfo {author} {\bibfnamefont {C.}~\bibnamefont
  {Sirtoti}},\ }\href@noop {} {\bibfield  {journal} {\bibinfo  {journal} {Phys.
  Rev. B}\ }\textbf {\bibinfo {volume} {85}},\ \bibinfo {pages} {045304}
  (\bibinfo {year} {2012})}\BibitemShut {NoStop}%
\end{thebibliography}%

\end{document}